\documentclass[preprint,12pt,authoryear]{elsarticle}

\usepackage{graphicx} 

\usepackage{amssymb}
\usepackage{amsmath}
\usepackage{nicefrac}

\usepackage{multirow, makecell, booktabs}

\usepackage[colorlinks=true, allcolors=blue]{hyperref}
\usepackage{cleveref}
\usepackage{todonotes}
\usepackage{subcaption}

\usepackage{xspace} 

\newcommand{\f}{\text{f}}
\newcommand{\s}{\text{s}}
\newcommand{\DuMuX}{DuMu$^\textrm{x}$\xspace}

\journal{International Journal of Heat and Mass Transfer}

\begin{document}

\begin{frontmatter}



\title{Local Thermal Non-Equilibrium Models in Porous Media: A Comparative Study of Conduction Effects} 

\author[iws]{Anna Mareike Kostelecky\corref{cor1}}
\ead{annamareike.kostelecky@iws.uni-stuttgart.de}
\author[uib,vista]{Ivar Stefansson} 
\author[hvl]{Carina Bringedal}
\author[iws]{Tufan Ghosh}
\author[uib]{Helge K. Dahle}
\author[iws]{Rainer Helmig}

 \cortext[cor1]{Corresponding author}

\affiliation[iws]{organization={Institute for Modelling Hydraulic and Environmental Systems, Univeristy of Stuttgart},
            city={Stuttgart},
            country={Germany}}
\affiliation[uib]{organization={Department of Mathematics, University of Bergen},
            city={Bergen},
            country={Norway}}
\affiliation[vista]{organization={Center for Modeling of Coupled Subsurface Dynamics, Department of Mathematics, University of Bergen},
            city={Bergen},
            country={Norway}}
\affiliation[hvl]{organization={Department of Computer science, Electrical engineering and Mathematical sciences, Western Norway University of Applied Sciences},
            city={Bergen},
            country={Norway}}

\begin{abstract}
Instantaneous heat transfer between different phases is a common assumption for modeling heat transfer in porous media, known as Local Thermal Equilibrium (LTE).
This assumption may not hold in certain technical and environmental applications, especially in systems with large temperature gradients, large differences in thermal properties, or high velocities.
Local Thermal Non-Equilibrium (LTNE) models aim to describe heat transfer processes when the LTE assumption may fail.

In this work, we compare three continuum-scale models from the pore to the representative elementary volume (REV) scale. Specifically, dual-network and REV-scale models are evaluated against a pore-resolved model, which we perceive as a reference in the absence of experimental results. Different effective models are used to obtain upscaled properties on the REV scale and to compare resulting temperature profiles.

The systems investigated are fully saturated, consisting of one fluid and one solid phase.
This study focuses on purely conductive systems without significant differences in thermal properties. Results show that LTE holds then for low interfacial resistances. However, for large interfacial resistances, solid and fluid temperatures differ.
The REV-scale model with effective parameters obtained by homogenization leads to similar results as the pore-resolved model, whereas the dual-network model shows greater deviation due to its fixed spatial resolution. Among the evaluated effective parameter formulations for the REV-scale model, only the homogenization-based approach captures the LTNE behavior, as it incorporates the interfacial heat transfer coefficient.
Convection is relevant for most practical applications, and its impact will be addressed in a follow-up article.
\end{abstract}




\end{frontmatter}


\section{Introduction}
    A common assumption in modeling thermal processes of porous medium systems is that the system is in Local Thermal Equilibrium (LTE), meaning instantaneous heat transfer between the bulk phases.
    However, for applications with large temperature gradients, fast dynamics or large differences in thermal properties, LTE may not be a valid assumption. In such cases, 
    models accounting for Local Thermal Non-Equilibrium (LTNE) effects should be introduced. LTNE models can then also be used to investigate the applicability of the LTE assumption. 
    There is a wide range of technical and environmental applications for which the LTE assumption may fail and LTNE models are required. These applications range from geothermal energy production \citep{gelet_2012_thermohydromechanical}, heat and water management in fuel cells \citep{hwang2006heat}, subsurface remediation through steam injection \citep{xu2023understanding}, self-pumping transpiration cooling \citep{dahmen2014numerical}, drying of thin porous media for industrial applications \citep{mujumdar2006handbook}, and evaporation at the interface between the subsurface and the atmosphere \citep{shahraeeni2010thermo}, among others.
    
    In this article, we focus on porous media systems. In the context of LTE processes, it is assumed that the temperatures of the respective phases equilibrate immediately, for example within a given control volume or at a certain position, such as the distinct interface between the phases. 
    Consequently, it is sufficient to consider only one temperature at the location of the interface or within the control volume. However, if the assumption of LTE does not apply, different temperatures of the respective bulk phases (e.g., one fluid and one solid temperature) as well as the interfacial heat transfer between the phases must be considered.
    
    The range of models developed to describe LTNE processes extends from the molecular to the representative elementary volume (REV) scale. 
    On the molecular scale, non-equilibrium molecular dynamics is a common approach employed to model processes with the objective of obtaining material or transport properties of a given system. This includes the Kapitza resistance of surfaces (see, e.g., \citet{gonccalves2022interfacial}). Additionally, various Lattice Boltzmann methods incorporate LTNE effects for porous media systems. Although Lattice Boltzmann methods are primarily applied to investigate systems at the pore scale, certain approaches also extend this to the REV scale (see, e.g., \citet{yang2020multiple}).
    LTNE models for porous media that are based on continuum theory are used at the pore and the REV scale. At the pore scale, this includes models that resolve the underlying geometry and heat transport across interfaces, such as pore-resolved models, as well as models that approximate the underlying porous structure. The latter can, e.g., be obtained using idealized shapes that form a network such as pore-network models. To account for interfacial heat transport at the pore scale, a dual-network model  \citep{koch_dual_2021} has been developed. The dual-network model is constructed through two interconnected networks - one for the fluid and one for the solid phase. 
    At the REV scale, several models that describe the heat transfer processes in an averaged sense, are available. These models capture the effective behavior of the porous medium (see, e.g., \citet{nuske2015modeling}). Both the dual-network and the REV-scale models are designed to handle systems at a larger scale than pore-resolved models. This comes at the cost of more model assumptions. Nevertheless, dual-network models can give insight into pore-scale processes without resolving the underlying porous geometry. REV-scale models are the basis for investigating large-scale systems. However, they require reliable approximations for effective parameters. This may be very challenging to obtain experimentally due to the limited ability to resolve local processes. Homogenization approaches can hereby help to provide these effective parameters for REV-scale models.
    
    Previous works, such as \citet{quintard1993one} and \citet{pati2022critical} give an overview of modeling approaches across different scales and the applicability of one- and two-equation continuum-scale approaches. The work of \citet{quintard1993one} focuses on conductive processes and performs volume averaging to obtain upscaled models. \citet{pati2022critical} review LTE and LTNE models additionally in case of convection and also provide an overview of research related to entropy balance. In case of LTE, \citet{aichlmayr_effective_2006} present a comparison of different effective thermal conductivity models for porous media systems filled with a single fluid phase.

    This study mainly addresses three aspects. First, we compare a dual-network and an REV-scale model against a pore-resolved reference model to assess their ability to capture LTNE effects. Secondly, our objective is to investigate the difference between LTE and LTNE models for scenarios of varying interfacial thermal resistances. Thirdly, we evaluate three different approximations for effective REV-scale parameters at LTNE conditions.
    Throughout this work, we only consider fully-saturated systems consisting of one fluid and one solid phase.
    To simplify the system and isolate different effects of heat transport, we focus solely on pure heat conduction, neglecting energy transport through convection and mass transport.
    
    The subsequent sections of the article are structured as follows: In \Cref{sec:models}, the three underlying mathematical model concepts along with their numerical simulation frameworks are introduced. Next, in \Cref{sec:comparison}, the setup of the comparison study between the three models is presented, followed by the results and their discussion. Finally, in \Cref{sec:finalRemarks}, we give concluding remarks of the study, including an outlook on future investigations. In \ref{app:homogenization_nonEq} to \ref{app:coarseREV}, we provide additional information on how specific model parameters are obtained, convergence results of the different models, as well as further investigations of the chosen setup.

\newpage
\section{Models} \label{sec:models}
\begin{figure}[h!]
        \centering
        \includegraphics[width=\linewidth]{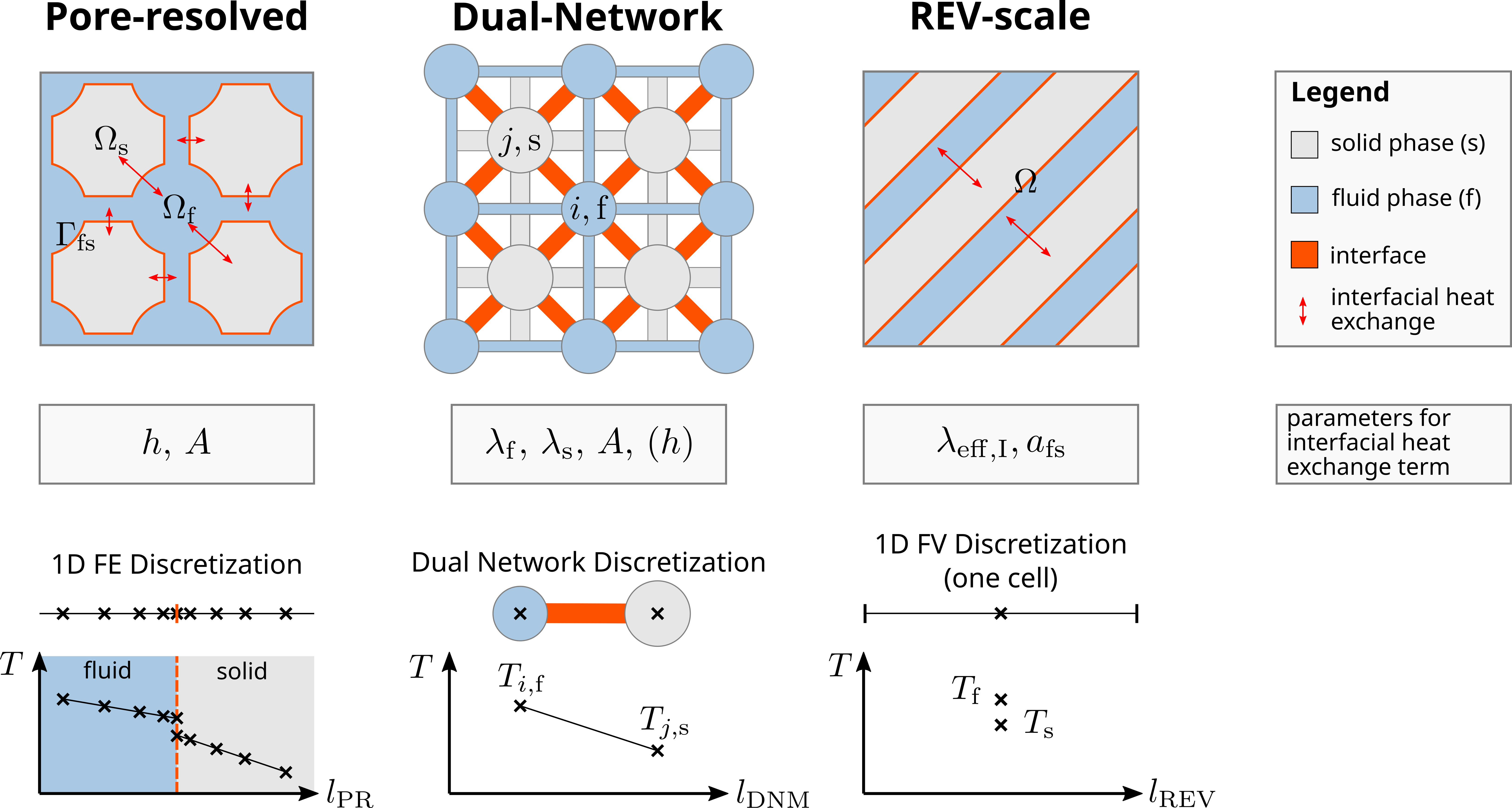}
        \caption{Overview of the different continuum-scale modeling approaches for taking LTNE into account. The first row presents the schematic figures of the models, the second row the parameters considered for the interfacial energy exchange, which are different from model to model. The last row shows the interpretation of the resulting discrete temperatures.}
        \label{fig:overview_models}
    \end{figure}

    This section introduces three classes of continuum-scale models that aim to describe heat transfer processes in porous media on different scales. We focus on LTNE processes, but will also state analogous LTE models for comparison. 
    The three model classes that will be compared in \Cref{sec:comparison} include pore-resolved, dual-network and REV-scale models. The pore-resolved model solves pore-scale balance equations for each phase on a grid that captures the pore-scale geometry of the system. Hereby, the sharp interface $\Gamma_{\f\s}$ between the void space, $\Omega_{\f}$, and the solid region, $\Omega_{\s}$, is resolved.
    Note that these two regions of our pore-scale domain, $\Omega_{\f}$ and $\Omega_{\s}$, are non-overlapping.
    An example of the pore-scale geometry is given in the upper row for the pore-resolved model of \Cref{fig:overview_models}. 
    The dual network consists of a void network that is coupled via interfacial connections to a solid network. The void-solid interfaces are not resolved in the dual-network model, though they still account for the pore-scale geometry in a simplified way.  In fact, the pore-scale geometry is reflected in the positions and parameters like the volume of the pore and solid bodies. These are indexed with $i,\f$ and $j,\s$ and depicted as blue and gray circles in \Cref{fig:overview_models}. In this model, balance equations are averaged over small void or solid volumes, i.e. pores and grains. 
    The REV-scale model is based on averaged balance equations over so-called representative elementary volumes (REV) within the domain $\Omega$. Hereby, the sharp interface between the phases or the pore-scale geometry will not be resolved. For meaningful averages, the REV length scale must be large enough for statistical homogeneity \citep{dullien2012porous}. Consequently, the two pore-scale models must cover multiple pores for comparability. However, practical comparisons are constrained by the computational efficiency of the pore-scale models.
    
    Each model describes the interfacial heat transfer in a distinct manner, resulting in different physical properties involved. The interfacial properties, including the interfacial area, $A$, and the heat transfer coefficient, $h$, are used in the pore-resolved model. The dual network additionally accounts for the thermal conductivities, $\lambda_{\f}$ and $\lambda_{\s}$, as the interface is not directly resolved. On the REV scale, upscaled properties such as the upscaled interfacial area $a_{\f\s}$ and an effective interfacial conductivity $\lambda_{\mathrm{eff,I}}$ are introduced. An overview of these parameters is provided in the second row of \Cref{fig:overview_models}.
    
    Since the three model approaches use different numerical schemes and resolve the processes in different detail, the resulting temperatures must be interpreted accordingly.
    The pore-resolved model, discretized through a finite element (FE) scheme, leads to fluid and solid temperatures at separate locations within $\Omega_{\f}$ and $\Omega_{\s}$, except at the interface $\Gamma_{\f\s}$. The dual network model, which employs a control volume finite element (CVFE) method, associates temperatures with the respective bodies without yielding an interface temperature. For the REV-scale model, discretized with a finite volume (FV) scheme, fluid and solid temperatures overlap within each control volume. This interpretation is illustrated in the last row of \Cref{fig:overview_models} for a one-dimensional case. Note that the length scales are different for the different model classes.
    Details of the models with their respective parameters as well as numerical simulation frameworks are given hereafter.
    
\subsection{General energy equations}
    For two inert phases, namely solid (s) and fluid (f), we assume that energy can only be transported by conduction. 
    In this case, the energy balance for each phase $\alpha$ can be formulated in terms of the phase temperature, $T_{\alpha}$, specific heat capacity, $c_{\alpha}$, the thermal conductivity, $\lambda_{\alpha}$, and the temperature $T_{\alpha}$ to
    \begin{equation}
        \partial_t\left(\rho_{\alpha} c_{\alpha} T_{\alpha} \right)
        - \nabla \cdot \left( \lambda_{\alpha} \nabla T_{\alpha}\right) = q_{\alpha}^e  \quad \text{in } \Omega_{\alpha} \,, \, \alpha = \{\f,\s\} \; .
        \label{eq:general_energy}
    \end{equation}
    For both energy equations, the sources and sinks of energy are expressed by the term on the right-hand side, $q_{\alpha}^e$.

    Based on these energy equations, different models for local equilibrium and local non-equilibrium processes in porous media can be developed depending on how the geometry may be simplified, and how the energy exchange between the solid and the fluid phase is modeled. 

    Note that in this study, and hence also in \Cref{eq:general_energy}, we neglect non-equilibrium thermal effects due to high velocities and therefore consider the phases to be at rest.
    
\subsection{Pore-resolved model} \label{sec:model_pore}
    The pore-resolved model aims to solve the energy equations in \Cref{eq:general_energy} without imposing geometrical simplifications of the pore-scale geometry. This means that we have separate energy equations for the respective domains $\Omega_{\f}$ and $\Omega_{\s}$ along with coupling conditions at the fluid-solid interface, $\Gamma_{\f\s}$. Whether the model describe LTE or LTNE processes therefore depends on how temperatures are coupled across internal interfaces $\Gamma_{\f\s}$.
    
\subsubsection{Pore-scale thermal equilibrium}\label{sec:model_pore_equil}
The equations in \eqref{eq:general_energy} are said to be in thermal equilibrium at the pore scale if the fluid phase and solid phase share the same temperature at their common interface $\Gamma_{\f\s}$. In this case, the appropriate internal interface conditions are
\begin{align}
	\lambda_{\f}\nabla T_{\f}\cdot \mathbf n &= \lambda_{\s}\nabla T_{\s}\cdot \mathbf n && \quad \text{on } \Gamma_{\f\s} \, , \label{eq:equil_flux} \\ 
	T_{\f} &= T_{\s} && \quad \text{on } \Gamma_{\f\s} \; .\label{eq:equil_T}
\end{align}
Here, $\mathbf n$ is the normal vector on the fluid-solid interface pointing into the fluid.

\subsubsection{Pore-scale thermal non-equilibrium}\label{sec:model_pore_nonequil}
If the fluid phase and solid phase do not have the same temperature at their common interface, the following internal interface conditions are used:
\begin{align}
	\lambda_{\f}\nabla T_{\f}\cdot \mathbf n &= \lambda_{\s}\nabla T_{\s}\cdot \mathbf n && \quad \text{on } \Gamma_{\f\s} \, ,\label{eq:nonequil_flux} \\ 
	\lambda_{\f}\nabla T_{\f}\cdot\mathbf n &= h(T_{\s}-T_{\f}) && \quad \text{on } \Gamma_{\f\s}  \; . \label{eq:nonequil_Tjump}
\end{align}
Here, $h$ represents the heat transfer coefficient of the interface depending on the two adjacent materials and has units of thermal conductivity divided by a length scale, $\frac{\text{W}}{\text{m}^2\text{K}}$. 
Note that \eqref{eq:nonequil_Tjump} reduces to \eqref{eq:equil_T} when $h\to\infty$ (or $1/h\to 0$). The presence of (a finite) $h$ means there is a contact resistivity between the fluid and solid. Locally at the interface, the temperatures will therefore not match if there is a temperature gradient in the system.
Also note that since $\lambda_{\f}\nabla T_{\f}\cdot \mathbf n = \lambda_{\s}\nabla T_{\s}\cdot \mathbf n$, the last condition can also be written $\lambda_{\s}\nabla T_{\s}\cdot\mathbf n = h(T_{\s}-T_{\f})$. 

For the comparison in \Cref{sec:comparison}, only the model with internal boundary conditions \eqref{eq:nonequil_flux} and \eqref{eq:nonequil_Tjump} are used, since LTE is covered in the limit $h\to\infty$. However, both models - LTE and LTNE - are taken to derive effective quantities for REV-scale models through homogenization (cf. \ref{app:homogenization_nonEq}).

\subsubsection{Simulation framework}\label{sec:model_pore_solve}
The pore-resolved model equations are discretized using a finite element approach. The domains $\Omega_{\f}$ and $\Omega_{\s}$ are gridded using Netgen \citep{schoberl1997netgen}, while the linear system of equations is assembled and solved using NGSolve \citep{schoberl2014c}. A relatively coarse mesh is used, since the domain of the system, $\Omega_{\f} \cup \Omega_{\s}$, must be large enough to be of the order of an REV.
 Note that Netgen needs the \emph{maximum} grid size to be specified. However, much smaller element sizes are used to resolve the specified domains, in particular near boundaries and interfaces. First order basis functions are used in the finite element approach. Time is discretized using implicit Euler, and a sparse Cholesky decomposition is used to find the inverse matrix used for the time stepping. For our simulations, we specify the maximum grid size $\Delta x_\text{max}=0.1/2800~\mathrm{m}$ and a fixed time-step size $\Delta t=0.1$. See \ref{app:conv_pore} for a convergence study. 
 The code for the pore-resolved simulations is available in \citet{bringedal2025code}.


\subsection{Dual-network model}
    
    The dual-network model, introduced by \citet{koch_dual_2021}, aims to describe mass and especially heat transfer processes in saturated porous media at the pore scale. For this, the pore-scale geometry is considered in a simplified manner. Two fully implicit network models \citep{weishaupt_fully_2022}, one for the pore space and one for the solid region of the porous medium, are monolithically coupled to account for heat transfer between them. The pore space is thus segmented into pore bodies and pore throats, while the solid region is divided into solid bodies and solid throats, representing the contact areas of the grains. Pore and solid bodies as well as throats are modeled through ideal shapes, such as spheres for the bodies and cylinders for the throats. 
\subsubsection{Model equations}    
    For the pore network, energy is stored in the pore bodies, whereas the pore throats allow heat transfer between the pore bodies. For a resting fluid phase $(\f)$ occupying the pore space, the energy balance for the pore body $(i,\f)$ becomes:
    \begin{equation}
        V_{i,\f} \, \partial_t \left(\rho_{\f} c_{\f} T_{i,\f}\right) -\sum_{i,\f \to j,\f} t_{ij,\f} \left(T_{i,\f} -T_{j,\f}\right) = - \sum_{{i,\f \to j,\s}} Q_{i,\f \to j,\s}  + Q_{i,\f}^e\, , \label{eq:DNM_f_energy_eq}
    \end{equation}
    where $V$ denotes the volume, $\rho$ the density, $c$ the specific heat capacity, $T$ the temperature and $t_{ij,\f}$ the thermal transmissibility of the pore throat $(ij,\f)$ connecting the pore bodies $(i,\f)$ and $(j,\f)$. The term $Q_{i,\f \to j,\s}$ represents the energy exchange between a pore body $(i,\f)$ and a solid body $(j,\s)$ and $Q_{i,\f}^e$ the source or sink term of the respective pore body. 
    
    Similarly, for the solid network the energy is stored within the so-called solid bodies, and the energy is transferred at the contact areas between those bodies, modeled by solid throats:
    \begin{equation}
        V_{j,\s} \,  \partial_t \left(\rho_{\s} c_{\s} T_{j,\s}\right) -\sum_{j,\s \to i,\s} t_{ij,\s} \left(T_{j,\s} -T_{i,\s}\right) = \sum_{{i,\f \to j,\s}} Q_{i,\f \to j,\s} + Q_{j,\s}^e
        \label{eq:DNM_s_energy_eq}
    \end{equation}
    with the same notation of physical quantities as for the pore network.
    
    \paragraph{Connections between pore and solid networks}
    Interfacial throats, connecting one pore body $(i,\f)$ to one solid body $(j,\s)$ or vice versa, allow for interfacial energy exchange through heat conduction
    \begin{equation}
        Q_{i,\f \to j,\s} = t_{i,\f \to j,\s}^{\mathrm{cond.}} \left(T_{i,\f} -T_{j,\s}\right) \; .
    \end{equation}
    Here, $t_{i,\f \to j,\s}^{\mathrm{cond.}}$ is the interfacial thermal transmissibility of the throat in case of a resting fluid.

    \paragraph{Thermal transmissibilities}
    The thermal transmissibilities incorporate the area and the length over which the energy is conducted, as well as a mean thermal conductivity $\bar{\lambda}$. Within the bulk phases $(\f)$ and $(\s)$, the transmissibility $t_{ij}$ is obtained by assuming a linear decrease of the area from the bodies to the center of the throat. Moreover, for the area of pore and solid bodies the effect of reduced space availability due to interfacial heat transfer is included through an effective area of these bodies $\tilde{A}_{i}$ (for details, see \citet{koch_dual_2021}). Assuming that all pore bodies and all solid bodies are of the same size, this leads to the thermal transmissibilities
    \begin{align}
        &t_{ij,\f} \approx \frac{\lambda_{\f} \sqrt{\tilde{A}_{i,\f}A_{ij,\f}}}{\Delta x_{ij, \f}}  \, , \qquad
        t_{ij,\s} \approx \frac{\lambda_{\s} \sqrt{\tilde{A}_{i,\s}A_{ij,\s}}}{\Delta x_{ij, \s}} \, , \\ &\text{with } \tilde{A}_{i}=f(A_{ij}, \kappa = \nicefrac{\lambda_{\f}}{\lambda_{\s}},C_0, C_{\infty}) \, ,
    \end{align}
    with the distance of two pore or solid bodies $\Delta x_{ij}$, the cross-sectional area of a pore or solid throat $A_{ij}$ and shape parameters $C_0$ and $C_{\infty}$.
    For interfacial heat transport from fluid to solid or vice versa, the thermal transmissibility $t_{i,\f \to j,\s}^{\mathrm{cond.}}$ is expressed through
    \begin{equation}
        t_{i,\f \to j,\s}^{\mathrm{cond.}} = \frac{C_I A_{i,\f \to j,\s}}{\Delta x_{i,\f \to j,\s}} \overline{\lambda}_{\f\s} \, , \label{eq:t_cond_1}
    \end{equation}
    depending on an interfacial shape factor $C_I$ and the interfacial area $A_{i,\f \to j,\s}$. Note that for this formulation from \citet{koch_dual_2021}, the sharp interface between the fluid and solid is assumed to have zero interfacial thermal resistance. This means that temperature continuity between the different phases is assumed. Hence, the mean thermal conductivity of \Cref{eq:t_cond_1} can be chosen as a distance-weighted harmonic mean between the intrinsic thermal conductivities of the phases
    \begin{equation}
        \overline{\lambda}_{\f\s} = \frac{\Delta x_{\f} + \Delta x_{\s}}{\frac{\Delta x_{\f}}{\lambda_{\f}} + \frac{\Delta x_{\s}}{\lambda_{\s}}}\, ,
    \end{equation}
    where $\Delta x_{\f}$ and $\Delta x_{\s}$ denote the distances from the center of the respective pore or solid body to the center of the interfacial throat connecting the pore and solid bodies.
    For a finite heat transfer coefficient $h$, the interface resists heat transport, with Kapitza resistance $\tilde{r}_{\f\s}$ defined as its inverse ($\tilde{r}_{\f\s} = \frac{1}{h}$). A non-zero Kapitza resistance causes a temperature jump across the sharp interface, motivating the mean thermal conductivity to be extended to
    \begin{equation}
        \overline{\lambda}_{\f\s} = \frac{\Delta x_{\f} + \Delta x_{\s}}{\frac{\Delta x_{\f}}{\lambda_{\f}} + \tilde{r}_{\f\s} + \frac{\Delta x_{\s}}{\lambda_{\s}}}\; .
    \end{equation}
    The choice of the shape parameters, briefly mentioned in this section, are elaborated in \ref{sec:DNM_shape_parameters}.
    
    \subsubsection{Simulation framework} For the comparison study in \Cref{sec:comparison}, the dual-network model will be run using the open source software framework \DuMuX \citep{Dumux2020}, designed to solve coupled transport processes in porous media.
    For the implementation, a control-volume finite element (CVFE) method, in particular the box-scheme, is used to solve the monolithic coupled dual-network model.
    Having a pore-scale geometry, the representation of the dual network can be obtained by a segmentation algorithm or through analytical calculation of the respective network properties.  For very simple, homogenous and isotropic pore-scale geometries the respective network properties, such as location, volume, cross-sectional areas etc., can be computed analytically and consequently the network can be directly constructed from this information. However, for real-world or more complex geometries, a segmentation algorithm, e.g., by the porous media image analysis toolkit PoreSpy \citep{gostick2019porespy}, has to be applied to get the respective network representing the pore space. Since the domain in \Cref{sec:comparison} is homogeneous and isotropic, the network is constructed analytically. Details about the implementation of boundary conditions for the setup investigated in \Cref{sec:comparison}, are given in \ref{sec:DNM_boundary}. For time discretization, a first-order backward Euler scheme is used together with an adaptive time-stepping scheme. As the time-step size is chosen based on the convergence rate of the Newton method, only a maximum time step of $\Delta t_{\mathrm{max}}=2.5$s is specified. 
    A time convergence study for the problem investigated in \Cref{sec:comparison} is presented in \ref{sec:conv_time_dnm}.
    The code related to the dual-network model is provided and can be run following the instructions in \citet{kostelecky2025code}.
    
\subsection{REV-scale model}\label{sec:model_rev}
    Accounting explicitly for individual pores, as for the two prevoius models, is not feasible for processes on larger scales.
    To reduce computational cost, REV-scale models account for geometrical properties and transport processes in an averaged sense.
    
\subsubsection{Local Thermal Equilibrium} \label{sec:REV-LTE}
    For LTE processes, different phases have the same temperature $T$ within a given control volume. Therefore, only one total energy balance for the tempertaure $T$ is needed: 
	\begin{equation}
		\partial_t \left(\Phi\rho_{\f} c_{\f} T\right)
		+ \partial_t \left(\left(1-\Phi\right)\rho_{\s} c_{\s} T \right)
		- \nabla \cdot \left( \lambda_{\mathrm{eff,pm}}  \nabla T \right) 
		=  q^{\text{energy}}  \quad \text{in } \Omega \; .
		\label{eq:REV_lte_eq}
	\end{equation}
	The phase properties $\rho$ and $c$ denote the density and specific heat capacity of the fluid (subscript $\f$) or the solid phase (subscript $\s$), $q^{\text{energy}}$ is the volume-specific energy source and $\Omega$ the REV-scale domain. In addition, the following REV-scale properties are used: the porosity $\Phi$ and the effective thermal conductivity of all phases $\lambda_{\mathrm{eff,pm}}$.
	
	\paragraph{Effective thermal conductivities}
	The effective thermal conductivity accounts for the fact that all bulk phases, either fluid or solid, have different thermal conductivities, and that conduction for those phases takes place over different area fractions.
	Several approaches that aim to incorporate these effects, are available. One approach is to use the volume fraction weighted arithmetic mean (see, e.g., \cite[p.39]{nield2006convection})
	\begin{equation}
		 \lambda_{\mathrm{eff,pm}}  = \Phi \lambda_{\f} + \left(1-\Phi\right) \lambda_{\s} \; .
	\end{equation}
	This is the most classical choice. However, other mean values, such as a weighted harmonic mean or a weighted geometric mean are possible \citep{nield2006convection}.\\ 
	For all volume fraction weighted means, it is to be assumed that the porosity of a system (volumetric property) is a good approximation for the area-specific porosity \citep{nield2006convection} as conduction takes place over cross-sections.
	Alternatively, homogenization theory is used to obtain an effective thermal conductivity for the entire porous medium, $\lambda_{\mathrm{eff,pm}}$, from pore-resolved simulations.  For the homogenization, pore-scale equations will be upscaled for a representative cell, to derive expressions for the effective quantities (see, e.g., \ref{app:homogenization_nonEq} and \citet{auriault2010homogenization}). The homogenization approach is chosen in \Cref{sec:comparison} for comparing the different model classes.
    
\subsubsection{Local Thermal Non-Equilibrium} \label{sec:rev_ltne_models}
	Allowing different phase temperatures within a given control volume, one energy equation per phase including the energy exchange between the phases must be accounted for. For one fluid ($\f$) and one solid phase ($\s$), the respective energy balances can be formulated as follows: 
	\begin{align}
		\partial_t \left(\Phi\rho_{\f} c_{\f} T_{\f}\right)
		- \nabla \cdot \left( \lambda_{\mathrm{eff,f}}  \nabla T_{\f}\right) 
		&= q_{\text{cond,s} \leadsto \f} + q_{\f}^{\text{energy}}   &&\quad \text{in } \Omega \, , \label{eq:REV_ltne_eq_fluid} \\ %
		\partial_t \left((1-\Phi)\rho_{\s} c_{\s} T_{\s}\right)
		- \nabla \cdot \left( \lambda_{\mathrm{eff,s}} \nabla T_{\s}\right) 
		&= - q_{\text{cond,s} \leadsto \f} + q_{\s}^{\text{energy}}    &&\quad \text{in } \Omega \, ,
		\label{eq:REV_ltne_eq_solid}
	\end{align}
	where $T_{\f}$ and $T_{\s}$ denote the fluid and solid temperature in a control volume. Furthermore, $\lambda_{\mathrm{eff,f}}$ and $\lambda_{\mathrm{eff,s}}$ are the effective thermal conductivities of the respective phase accounting for the area fraction available for heat transport through conduction within each phase.

	The interfacial conductive energy exchange term,
	\begin{equation}
		q_{\text{cond,s} \leadsto \f} = \lambda_{\mathrm{eff,I}}
		a_{\text{fs}} \left(T_{\s} - T_{\f} \right) \, \label{eq:exchange_s_f}
	\end{equation}
	describes the heat transport from the solid to the fluid phase and incorporates the volume-specific interfacial area $a_{\f\s}$ between the two phases (units of $\nicefrac{\mathrm{m}^2}{\mathrm{m}^3}$).
	Note that the effective interfacial thermal conductivity $\lambda_{\mathrm{eff},I}$ is in units of a bulk conductivity divided by a length.
	
	\paragraph{Effective thermal conductivities}
	Similarly to the effective thermal conductivity in case of LTE, the effective thermal conductivity for each phase can be obtained through scaling the bulk phase conductivities with the volume fraction of each phase (cf. \citet{nield2006convection}, \citet{nuske2015modeling}):
	\begin{align}
		&\left[\lambda_{\mathrm{eff,f}}\right]^{\text{Nuske}} = \Phi \lambda_{\f} \, , \label{eq:rev_effLambdaF_classical}\\
		&\left[\lambda_{\mathrm{eff,s}}\right]^{\text{Nuske}} = (1-\Phi)\lambda_{\s} \; . \label{eq:rev_effLambdaS_classical}
	\end{align}
	 Another approach considering additionally the dependence on the  conductivity ratio $\kappa=\nicefrac{\lambda_{\f}}{\lambda_{\s}}$ between the phases, follows after \citet{nakayama_two-energy_2001}:
    \begin{align}
		&\left[\lambda_{\text{eff},\f}\right]^{\text{Nakayama}} = \left(\Phi + (1-\kappa) G\right) \lambda_{\f} \label{eq:rev_lambdaEffNakayama_f} \, , \\
		&\left[\lambda_{\text{eff},\s}\right]^{\text{Nakayama}} = \left(1-\Phi + (\kappa - 1) G\right) \lambda_{\s} \label{eq:rev_lambdaEffNakayama_s} \; .
	\end{align}
	Here, the function $G$ depends on the intrinsic thermal conductivities as well as the porosity. 
	However, energy transport within a phase will most likely be restricted by the smallest geometrical constriction, e.g., the contact area of two grains. As this can be much smaller than the averaged, volume-specific area available for heat transport, again effective thermal conductivities obtained through homogenization theory will be considered (see, e.g., \ref{app:homogenization_nonEq}).
	
	For the interfacial thermal conductivities, the formulation in \citet{nuske2015modeling} can be simplified for a stagnant fluid phase to
	\begin{equation}
		\left[\lambda_{\mathrm{eff,I}}\right]^{\text{Nuske}} =  \frac{ \overline{\lambda}_{\text{fs}} }{L_{\mathrm{ch}}} \, ,
        \label{eq:rev_effLambdaI_Nuske}
	\end{equation}
	where $\overline{\lambda}_{\text{fs}}$ denotes the harmonic mean of the intrinsic phase conductivities and $L_{\mathrm{ch}}$ is a characteristic length scale of the system. 
    This characteristic length scale is typically chosen as the average grain diameter for porous media systems \citep{bear1972dynamics}. We will use the diameter $d_{\mathrm{50}}$, defined as the diameter of a granular material below which $50\%$ of the grains fall by mass \citep{bear1972dynamics}.
	Homogenization of the pore-scale equations leads to (cf. \ref{app:homogenization_nonEq}):
	\begin{align}
		&\left[\lambda_{\mathrm{eff,I}} a_{\f\s}\right]^{\text{homogenization}} = H = h a_{\f\s} \, , \label{eq:rev_effLambdaI_homog}
	\end{align}
	where $h$ denotes the heat transfer coefficient as introduced in \Cref{sec:model_pore_nonequil}.
    The factor of interfacial heat exchange term (cf. \Cref{eq:exchange_s_f}) after \citet{nakayama_two-energy_2001}, reads
    \begin{equation}
        \left[\lambda_{\mathrm{eff,I}}a_{\f\s} \right]^{\text{Nakayama}} = a_{\f\s} h_{\s\f} - \lambda_{\s}G \nabla^2\; . 
    \end{equation}
    Here, $h_{\s\f}$ denotes an interfacial heat transfer coefficient obtained from an empirical formulation for densely packed beds, see \citet{wakao1982heat}, and $G$ denotes the tortuosity parameter, given through another empirical formulation in \citet{nakayama_two-energy_2001}.
    The first term corresponds to the previous two formulations (cf. \Cref{eq:rev_effLambdaI_Nuske} and \Cref{eq:rev_effLambdaI_homog}), the second term includes the Laplacian of the temperature difference, which increases the interfacial exchange term due to tortuosity.
    
    \subsubsection{Simulation Framework}
    For the REV model, we use a finite volume spatial discretization and an implicit Euler temporal discretization of the REV-scale equations implemented in the PorePy simulation toolbox \citep{keilegavlen2021porepy}.
    For the simple geometries and isotropic media considered herein, we employ the two-point flux approximation for the conductive flux, noting that a more advanced multi-point is also available in the toolbox.
    PorePy has a native model for advective and conductive heat transport under the LTE assumption.
    Using the framework introduced by \citet{stefansson2024flexible}, extension to the LTNE case is relatively straightforward. 
    These extensions are available in the run scripts at \citet{stefansson2025code}.
    
\newpage
\section{Comparison study} \label{sec:comparison}
\begin{figure}[ht!]
        \centering
        \includegraphics[width=0.85\linewidth]{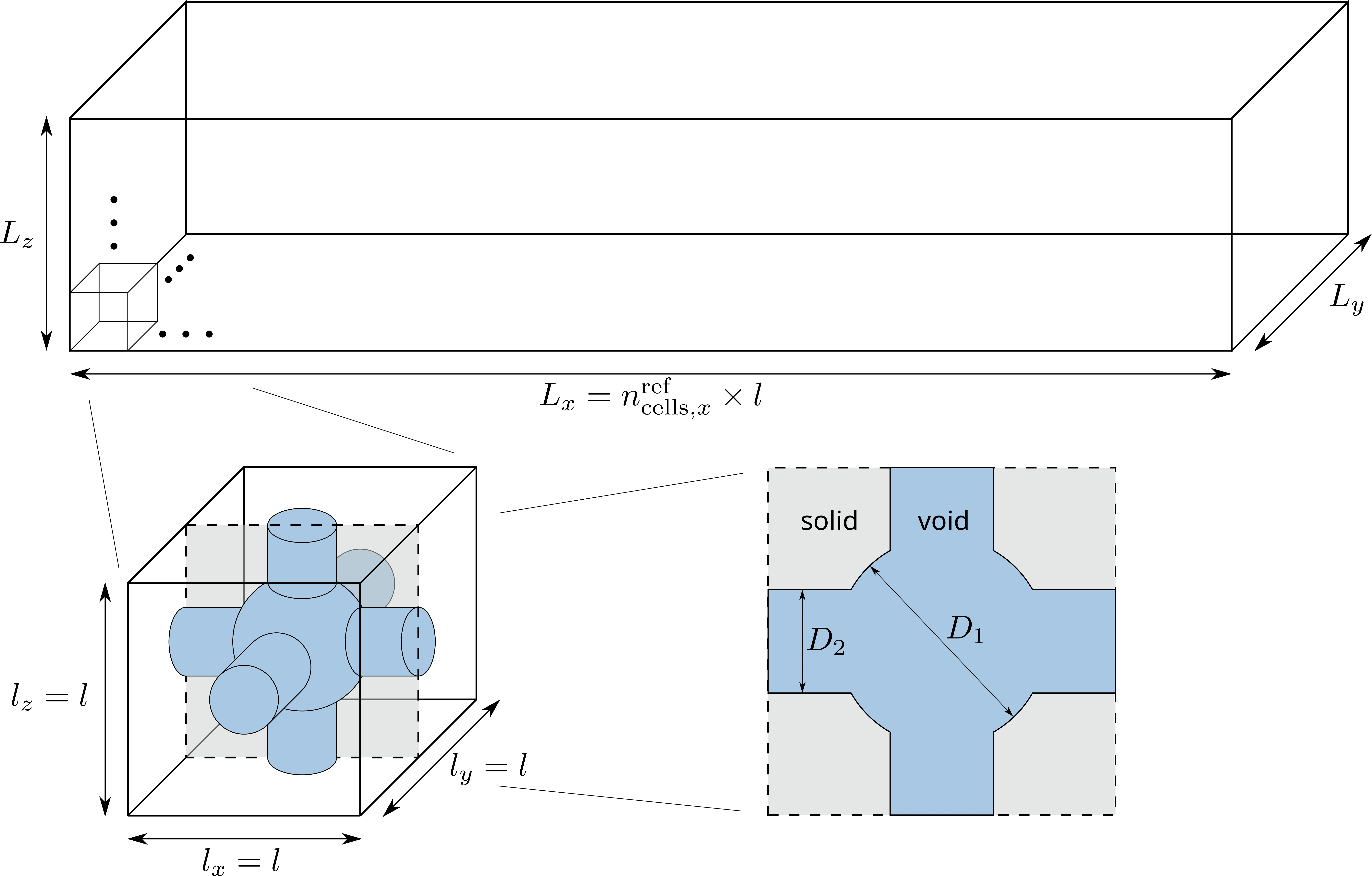}
        \caption{Geometrical setup of the simulation domain with $n_{\mathrm{cells},x}^{\mathrm{ref}}$, $n_{\mathrm{cells},y}^{\mathrm{ref}}$, $n_{\mathrm{cells},z}^{\mathrm{ref}}$ reference cells in the coordinate axis directions. One three-dimensional, cubic reference cell with the side lengths $l$ is shown, where six tubes - two in each direction - connected to one sphere. A cut through the three-dimensional reference cell visualizes the void space with the respective diameters of the sphere and the tubes, as well as the solid phase.}
        \label{fig:setup}
    \end{figure}

    To compare the three models introduced in \Cref{sec:models}, we consider the following setup as depicted in \Cref{fig:setup}.
	The computational domain is chosen to consist of $n_{\mathrm{cells},x}^{\mathrm{ref}} \times n_{\mathrm{cells},y}^{\mathrm{ref}} \times n_{\mathrm{cells},z}^{\mathrm{ref}}$ reference cells in the direction of the respective coordinate axis, where $n_{\mathrm{cells},x}^{\mathrm{ref}} = 45$ and $n_{\mathrm{cells},y}^{\mathrm{ref}} = n_{\mathrm{cells},z}^{\mathrm{ref}} = 4$. Each reference cell is identical and has a cubic shape with a side length of $l=\nicefrac{1}{2800} ~ \text{m}\approx 3.6\times 10^{-4} ~\mathrm{m}$.
    A reference cell is divided into two distinct regions: a void space and a space occupied by the solid phase. The void space is constructed of six tubes, each with a diameter of $D_2 = 10^{-4} ~ \mathrm{m}$, which are connected to a sphere with a diameter of $D_1 = 2.6\times 10^{-4} ~\mathrm{m}$.
	From this pore-scale geometry, we can analytically calculate REV-scale quantities such as porosity, $\Phi=0.25446$, and volume-specific interfacial area, $a_{\mathrm{fs}} = 6511.1503 ~\frac{\mathrm{m}^2}{\mathrm{m}^3}$.
    For the solid phase we choose the material properties of granite, while for the liquid phase occupying the void space, water is taken. Note that the respective material properties, presented in \Cref{tbl:material_properties}, are assumed to be constant in space and time and do not change with e.g., temperature, pressure, etc.
    The simulation data as well as the post-processing routine can be found in \citet{kostelecky2025postprocessing}.

    \begin{table}[h!]
		\centering
		\begin{tabular}{ccc}
			\toprule
            \textbf{Property} & \textbf{Fluid} & \textbf{Solid} \\ \midrule
            Density $\left[\frac{\mathrm{kg}}{\mathrm{m}^3}\right]$& $10^3$ & $2.7 \times 10^3$\\ \addlinespace
            Heat capacity $\left[\frac{\mathrm{J}}{\mathrm{kgK}}\right]$& $4180$ & $790$ \\ \addlinespace
            Thermal conductivity $\left[\frac{\mathrm{W}}{\mathrm{mK}}\right]$& $0.679$ & $2.8$ \\
            \addlinespace
            \bottomrule
		\end{tabular}
		\caption{Material properties for the liquid and the solid phase.}
		\label{tbl:material_properties}
	\end{table}

    \paragraph{Boundary and initial conditions}
    Initially, we set the temperatures of both phases to $293.15~$K.
	For all models accounting for two energy equations, one solid and one fluid, we set the following boundary conditions.
    On the left boundary, the fluid temperature is increased by $10~$K with respect to the initial value. For the solid phase, isolating boundary conditions are enforced on the left side via a zero normal temperature gradient.
	On the remaining boundaries, zero-Neumann boundary conditions for the temperatures of both phases are applied. \Cref{tbl:Ics_Bcs} summarizes the initial and boundary conditions for the fluid and the solid phase.
    In case of the LTE-REV model (cf. \Cref{sec:REV-LTE}), where only one energy balance is solved for both phases, we set a Dirichlet boundary condition for the temperature with a value of $T_{\mathrm{ini}} + 10$ K on the left side, while enforcing isolating boundary conditions on all remaining boundaries.
    By the choice of the before-mentioned boundary conditions the test case is quasi-1d, leading to minimal dynamics in the $y-$ and $z-$direction.

	\begin{table}[h!]
		\centering
		\begin{tabular}{lccc}
			\toprule
			\multicolumn{2}{l}{\textbf{Conditions}}  &\textbf{Fluid} & \textbf{Solid} \\ \hline
			\multirow{3}{*}{\rotatebox[origin=c]{90}{Initial}} & & \multirowcell{3}{$T_{\mathrm{ini},\f}= T_{\mathrm{ini}} = 293.15 \, \mathrm{K}$} & \multirowcell{3}{$T_{\mathrm{ini},\s}= T_{\mathrm{ini}} = 293.15 \, \mathrm{K}$} \\
			& & & \\ 
			& & & \\ \hline
			\multirow{4}{*}{\rotatebox[origin=c]{90}{Boundary}} & \multirow{2}{*}{Left} & \multirowcell{2}{$T_{\mathrm{left},\f} = T_{\mathrm{ini}} + 10 \, \mathrm{K}$}& \multirowcell{2}{$\nabla T_{\s} \cdot \mathbf{n} = 0$}\\
			& &  & \\ \cline{2-4}
			& \multirow{2}{*}{Others} & \multirowcell{2}{$\nabla T_{\f} \cdot \mathbf{n} = 0$} & \multirowcell{2}{$\nabla T_{\s} \cdot \mathbf{n} = 0$}\\
			& &  & \\ \bottomrule
		\end{tabular}
		\caption{Initial and boundary conditions for fluid and solid phase.}
		\label{tbl:Ics_Bcs}
	\end{table}

    \paragraph{Evaluation of temperatures}
    For the comparison study between the different models, the averaged temperatures are compared at time $t_1=5~$s and $t_2=50~$s. In order to compare pore-scale results, obtained from the pore-resolved and the dual-network model, with the REV-scale results, the averaged temperatures are chosen as the mean values over one REV. Each REV is hereby chosen to consist of $15\times4\times4$ reference cells, which is considered to result in a sufficient number of pores to characterize an REV \citep{dullien2012porous}. The effective thermal conductivities for the REV-scale models, $\lambda_{\mathrm{eff},\f},\,\lambda_{\mathrm{eff},\s}$ and $\lambda_{\mathrm{eff,pm}}$, that are used in \Cref{sec:case1} and \Cref{sec:case2} are taken from homogenization (see \ref{app:homogenization_nonEq}). For the dual-network and REV-scale model, temperatures from LTNE as well as temperature continuity at the fluid-solid interface will be evaluated. For the latter case, the REV-scale model simplifies to the LTE model, leading to only one averaged temperature instead of one per bulk phase. For the pore-resolved model, only the model for LTNE processes is considered as this serves as a reference.
    
    \paragraph{Test cases}
    Sections \ref{sec:case1} and \ref{sec:case2} consider two different values of the heat transfer coefficient, with the aim of comparing the three classes of models in case of low and high interfacial resistivities. In \Cref{sec:case1}, an interfacial thermal resistance for water-silicon interfaces with $\tilde{r}_1=1.2\times 10^{-8} ~\frac{\text{m}^2\text{K}}{\text{W}}$  \citep{gonccalves2022interfacial} is used to mimic a water-granite system. For the second case, an interfacial resistance value of $\tilde{r}_2=10^{-2} ~\frac{\text{m}^2\text{K}}{\text{W}}$ is chosen. In this case, the thermal conductivities of the bulk phases are kept unchanged to solely investigate the effect of the interfacial resistance, without considering the coupled effect of the bulk phases and the interface between them. 
    Although the value does not correspond to the water-granite system described above, such high resistance values are observed, e.g., for copper-helium interfaces; see \citet{Pollack1969Kapitza}.
    The respective values for the heat transfer coefficients, $h_1 \approx 8.3\times10^7 ~\frac{\text{W}}{\text{m}^2\text{K}}$ and $h_2 = 100 ~\frac{\text{W}}{\text{m}^2\text{K}}$, follow from the inverse of the Kapitza resistances. 
	In \Cref{sec:comparison_REV}, the influence of different effective thermal conductivity formulations will be discussed.

\subsection{Comparison for a low interfacial resistance value (high heat transfer coefficient)} \label{sec:case1}

\begin{figure}
        \begin{subfigure}{\linewidth}
            \centering
            \includegraphics[width=\linewidth]{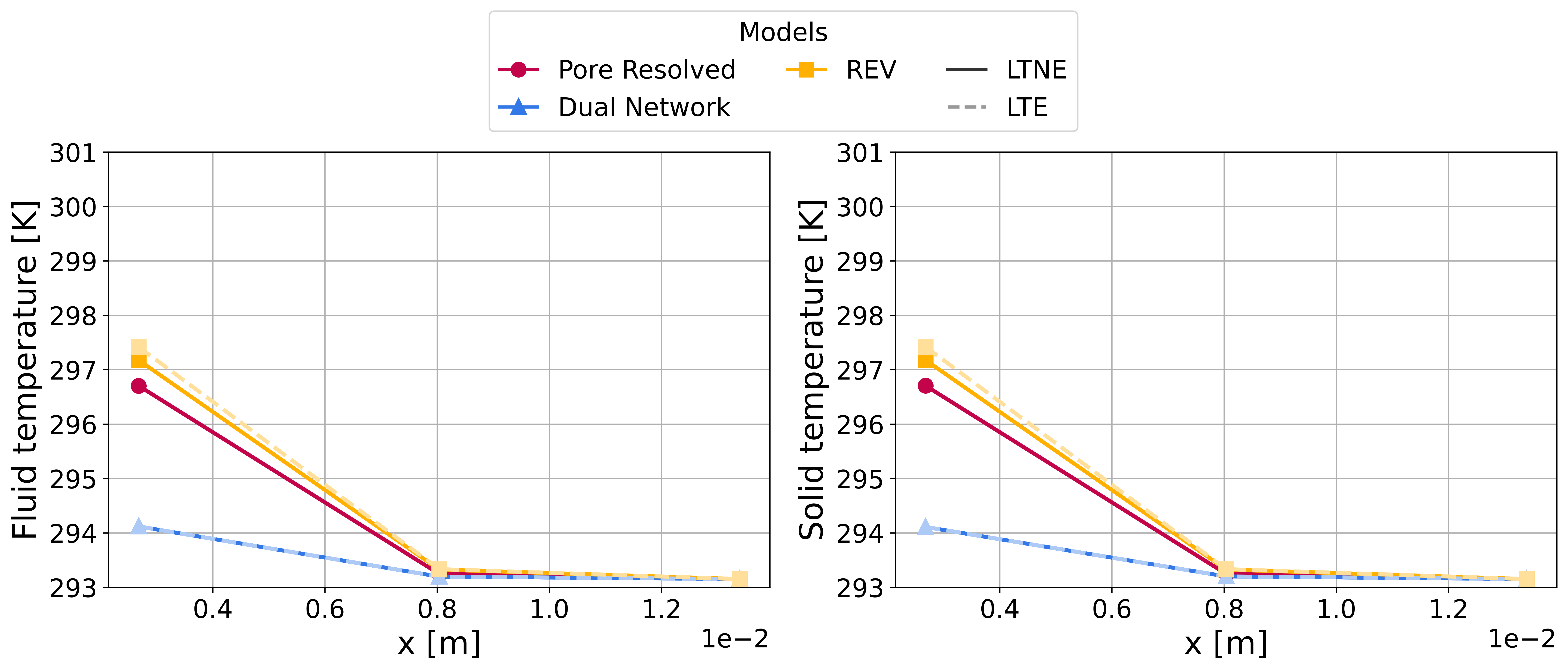}
            \caption{$t=5$s}
            \label{fig:case1_t1}
        \end{subfigure}
        \begin{subfigure}{\linewidth}
            \centering
            \includegraphics[width=\linewidth]{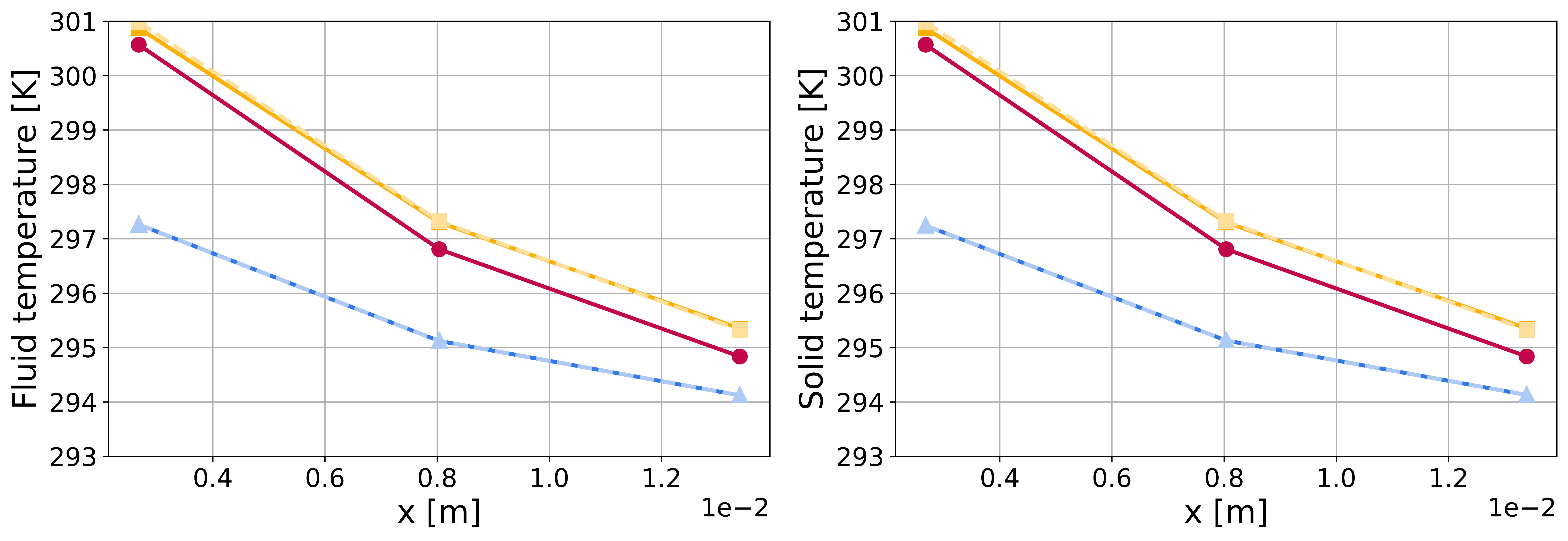}
            \caption{$t=50$s}
            \label{fig:case1_t2}
        \end{subfigure}
        \caption{Averaged REV-scale temperatures for three different model classes, which are distinguished by colors and markers. The LTNE concepts, where the interfacial heat transfer coefficient of $h_1 = 8.3\times10^7 ~\frac{\text{W}}{\text{m}^2\text{K}}$ is accounted for, are plotted in solid lines, while models neglecting this (LTE) are plotted in dashed lines and brighter colors. Only for the dual-network model and the REV-scale model, LTE model variants are used. Fluid temperatures are shown on the left side and solid temperatures on the right. The top row shows the evaluations at an earlier time $t_1$, while the bottom row shows them at a later time $t_2$.}
        \label{fig:case1}
    \end{figure}

    \Cref{fig:case1} shows the resulting averaged temperatures for all three model classes in case of a low interfacial thermal resistance, $\tilde{r}_1=1.2\times 10^{-8} ~\frac{\text{m}^2\text{K}}{\text{W}}$, correspondingly a high heat transfer coefficient of $h_1 \approx 8.3\times10^7 ~\frac{\text{W}}{\text{m}^2\text{K}}$. Converged results are used for both the dual-network and REV-scale models. The temperatures of the pore-resolved model are influenced by numerical diffusion due to using a relatively coarse mesh such that the domain of interest can be computed. The pore-resolved temperatures are therefore overpredicted, and we refer to \ref{app:convergence} for a convergence study of all models.
    
    First, we compare the fluid and solid temperature profiles between the same models or model classes.
    Comparing the temperatures of the two phases, it can be seen that for each model both temperatures are equal, and therefore the phases are in LTE. Hence, also the results for models considering the interfacial thermal resistance, denoted by LTNE, give (almost) the same results for the dual-network model and the REV-scale model compared to the results assuming temperature continuity at the underlying solid-fluid interface, denoted by LTE. For the dual-network model it can be shown that the LTNE formulation at the interface goes towards the LTE formulation as the heat transfer coefficient goes to infinity. 
    
    Secondly, the results of the dual-network and REV-scale model are compared to the pore-resolved model in case of the LTNE model version. Considering that the pore-resolved model is expected to overpredict the resulting temperatures, the pore-resolved and the REV-scale temperatures follow the same tendency in terms of the resulting temperature gradient within the system. However, the conduction process happens faster for the REV scale, which is due to the more diffusive nature of the model. 
    A larger difference can be observed for the dual-network model compared to the pore-resolved model, where the temperature profile has a less steep slope and averaged temperatures are lower. This can be explained by the fact that the spatial resolution of the dual-network model is fixed and therefore cannot be refined. A small comparison of the dual-network model results to results of REV-scale model simulations with a coarse spatial discretization of $n_{\mathrm{cells,x}}=45$ shows smaller differences between the dual-network and the REV-scale model (see \Cref{fig:case1_coarse} in \ref{app:coarseREV}), which confirms the influence of the spatial resolution. Thus, resolving the largest temperature differences that occur at the left boundary leads to more accurate results in case of the pore-resolved and REV models.
    
    Overall, the pore-resolved and REV-scale model feature a similar conductive timescale, indicating that the choice of the effective thermal conductivities for the REV-scale model are within a  reasonable range. However, the diffusive processes are slightly overestimated in comparison to pore-resolved results.
    An additional study for a coarse spatial discretization shows that differences between the dual-network model and the other two models are mainly addressed to the fixed spatial resolution. The thermal transmissibilities of the dual-network model can be considered plausible as shape parameters are chosen such that the effective conductivities obtained from the dual-network model are close to those of homogenization (see \ref{app:dnm_additional}). This choice leads then to similar temperatures when compared to results of a similar coarse REV-scale discretization.

\subsection{Comparison for a high interfacial resistance value (low heat transfer coefficient)} \label{sec:case2}

\begin{figure}
        \begin{subfigure}{\linewidth}
            \centering
            \includegraphics[width=\linewidth]{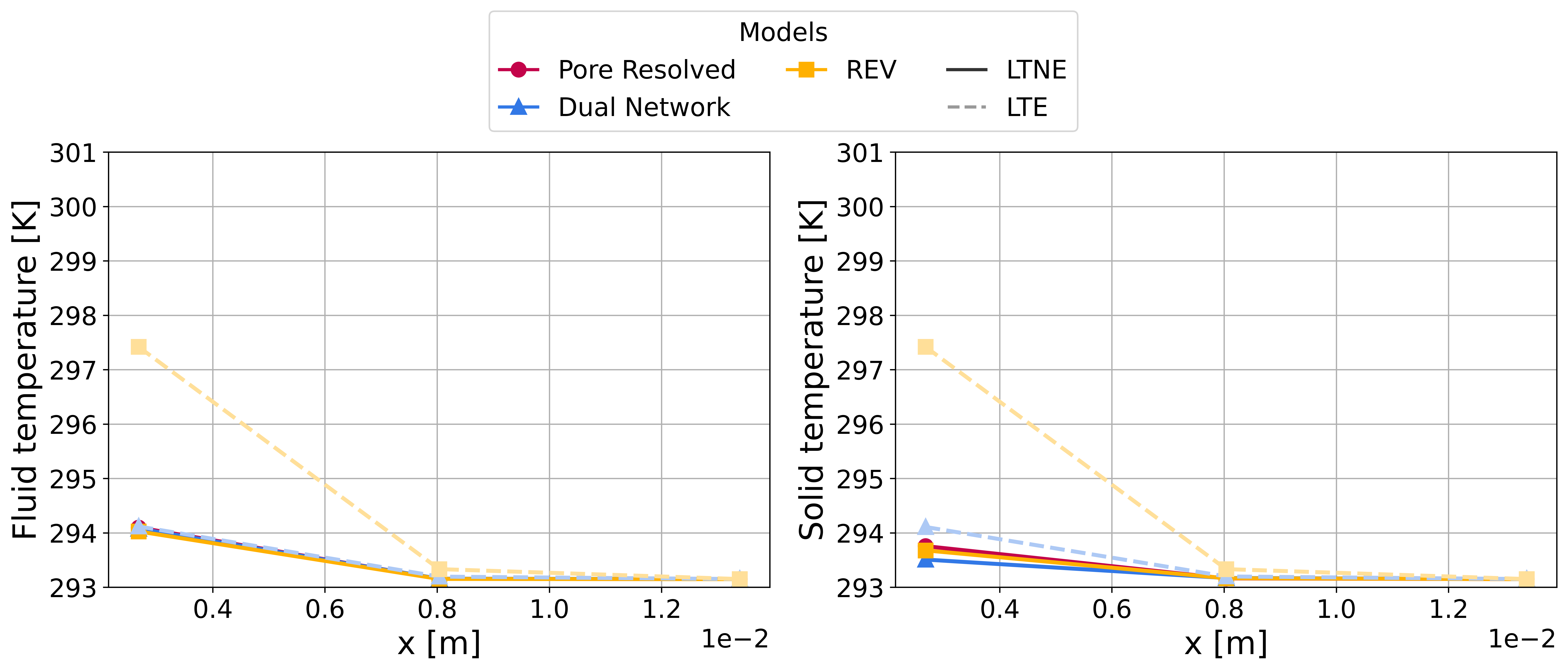}
            \caption{$t=5$s}
            \label{fig:case2_t1}
        \end{subfigure}
        \begin{subfigure}{\linewidth}
            \centering
            \includegraphics[width=\linewidth]{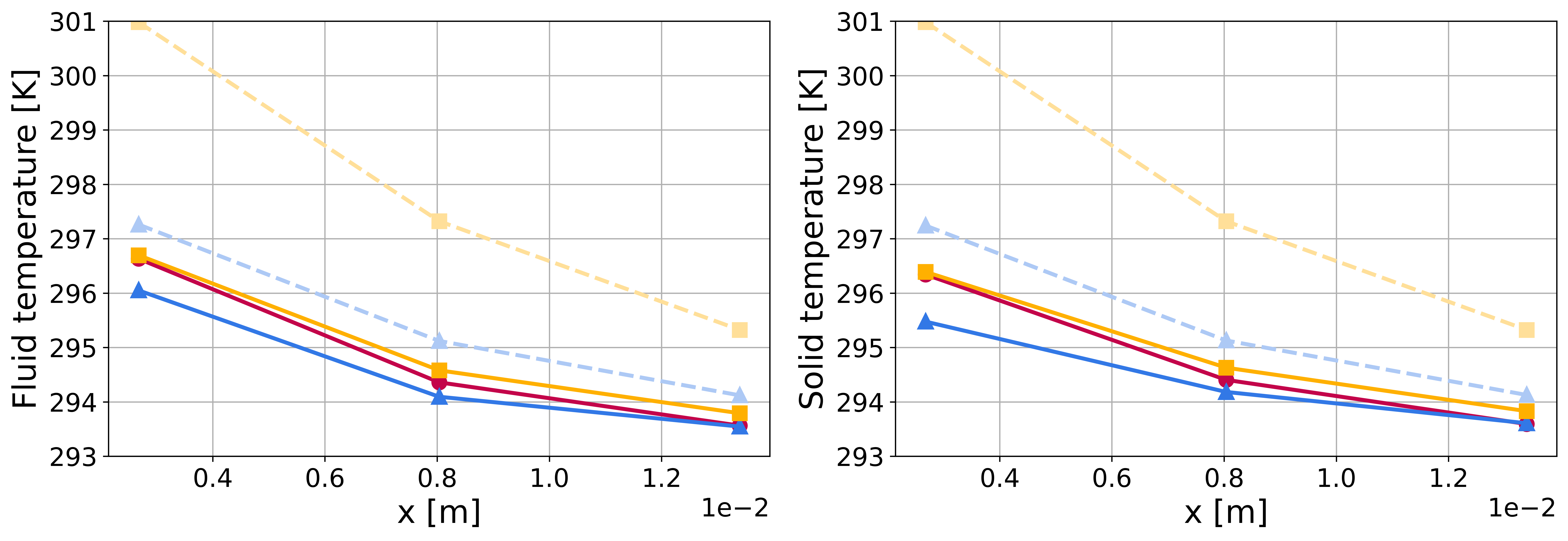}
            \caption{$t=50$s}
            \label{fig:case2_t2}
        \end{subfigure}
        \caption{Averaged REV-scale temperatures for three different model classes, which are distinguished by colors and markers. The LTNE concepts, where the interfacial heat transfer coefficient of $h_2 = 100 ~\frac{\text{W}}{\text{m}^2\text{K}}$ is accounted for are plotted in solid lines, while models neglecting this (LTE) are plotted in dashed lines and brighter colors. Only for the dual-network model and the REV-scale model, LTE model variants are used. Fluid temperatures are shown on the left side and solid temperatures on the right. The top row shows the evaluations at an earlier time $t_1$, while the bottom row shows them at a later time $t_2$.}
        \label{fig:case2}
    \end{figure}

    Effects of LTNE are expected to be relevant if the Biot number is small enough \citep{auriault2010homogenization}.
    The Biot number hereby relates the thermal conductivity of an interface with $h \times L$, where $L$ is the characteristic length of the system, to the thermal conductivity of a bulk phase (see also \ref{app:homogenization_nonEq}).
    Hence, to achieve a lower Biot number than used in \Cref{sec:case1}, either the value of the heat transfer coefficient, the characteristic length scale of the system or the phase thermal conductivity can be varied. We choose here a lower heat transfer coefficient with $h_2 = 100 ~\frac{\text{W}}{\text{m}^2\text{K}}$, while keeping the other values fixed.
    The resulting averaged temperatures for $h_2$, which corresponds to a Kapitza resistance of $\tilde{r}_2=10^{-2} ~\frac{\text{m}^2\text{K}}{\text{W}}$, are given in \Cref{fig:case2}.

    When the two resulting phase temperatures for each LTNE model are compared, it is clearly visible that the fluid and solid temperatures are no longer exactly the same. This is especially the case for the temperatures of the first REV (first evaluation from the left side). The average solid temperature on the left part of the domain does not instantly equilibrate to the corresponding higher fluid temperatures due to the added interfacial resistance. Moreover, a clear influence of the higher interfacial resistance is also visible when looking at the results in case of temperature continuity at the interface (LTE) in contrast to those when the heat transfer coefficient (LTNE) is accounted for. 
    The conduction process is happening noticeably slower when the low heat transfer coefficient and therefore the temperature jump across the fluid-solid interface is taken into account. This indicates the importance of the LTNE models to include the interfacial resistance into the formulations for cases of low heat transfer coefficients or, equivalently, small pore geometries.
    
    In contrast to the high interfacial heat transfer coefficient used in \Cref{sec:case1}, the absolute difference between the different LTNE models is smaller due to the slower dynamics of the system. Overall, the difference between the models is however similar as for \Cref{sec:case1}. 
    The REV-scale model with the effective conductivities from the homogenization theory leads to results very close to those of the pore-resolved models, but to a slightly faster process. The dual-network model again yields a less steep gradient of the average temperatures and slightly lower temperatures compared to the other two models originating from the fixed spatial discretization.
    \Cref{fig:case2_coarse} in \ref{app:coarseREV} shows  the comparison of the dual-network model and REV model with an equally coarse discretization as used for the network model.
    
\subsection{Investigation of different effective bulk phase conductivities for REV-scale models} \label{sec:comparison_REV}

\begin{figure}
        \begin{subfigure}{\linewidth}
            \centering
            \includegraphics[width=\linewidth]{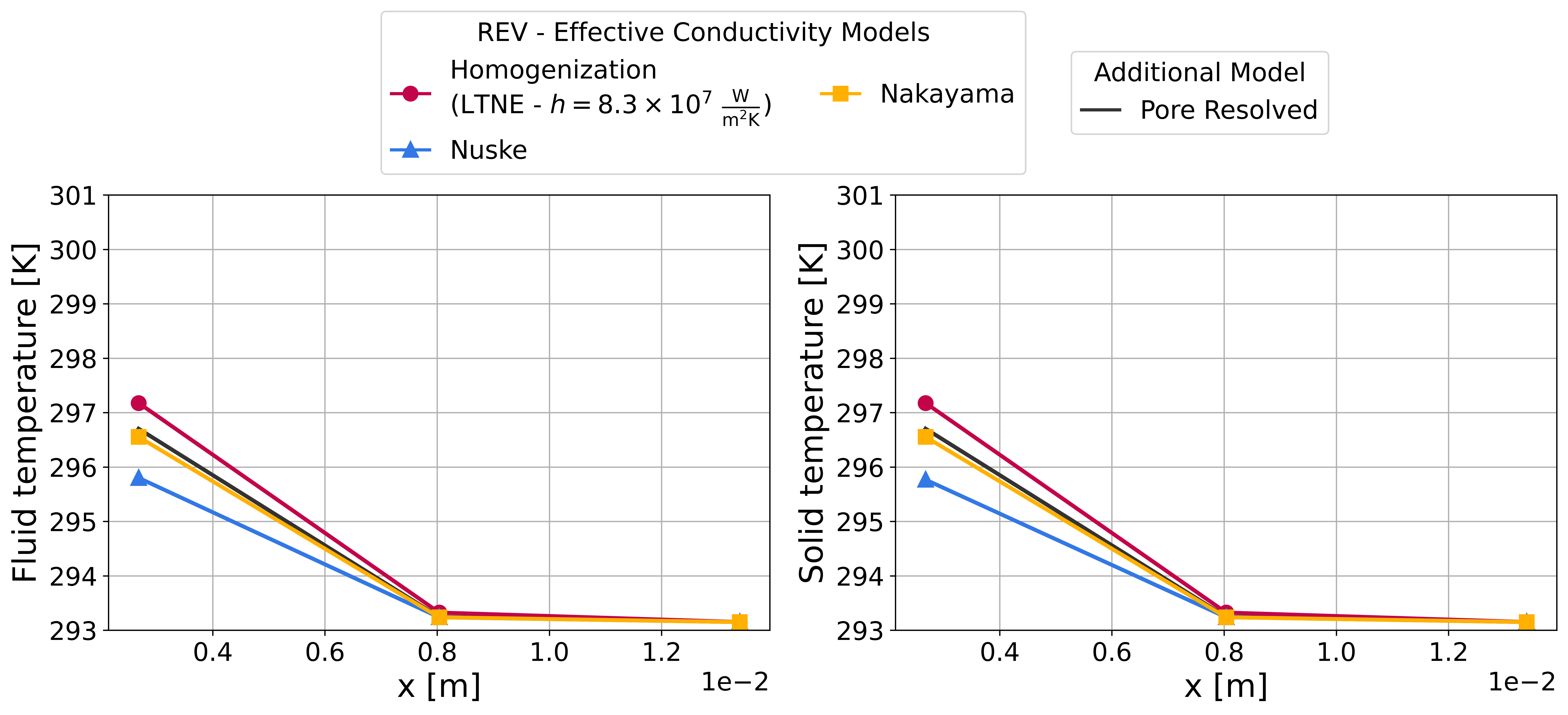}
            \caption{$t=5$s}
            \label{fig:rev_models_t1}
        \end{subfigure}
        \begin{subfigure}{\linewidth}
            \centering
            \includegraphics[width=\linewidth]{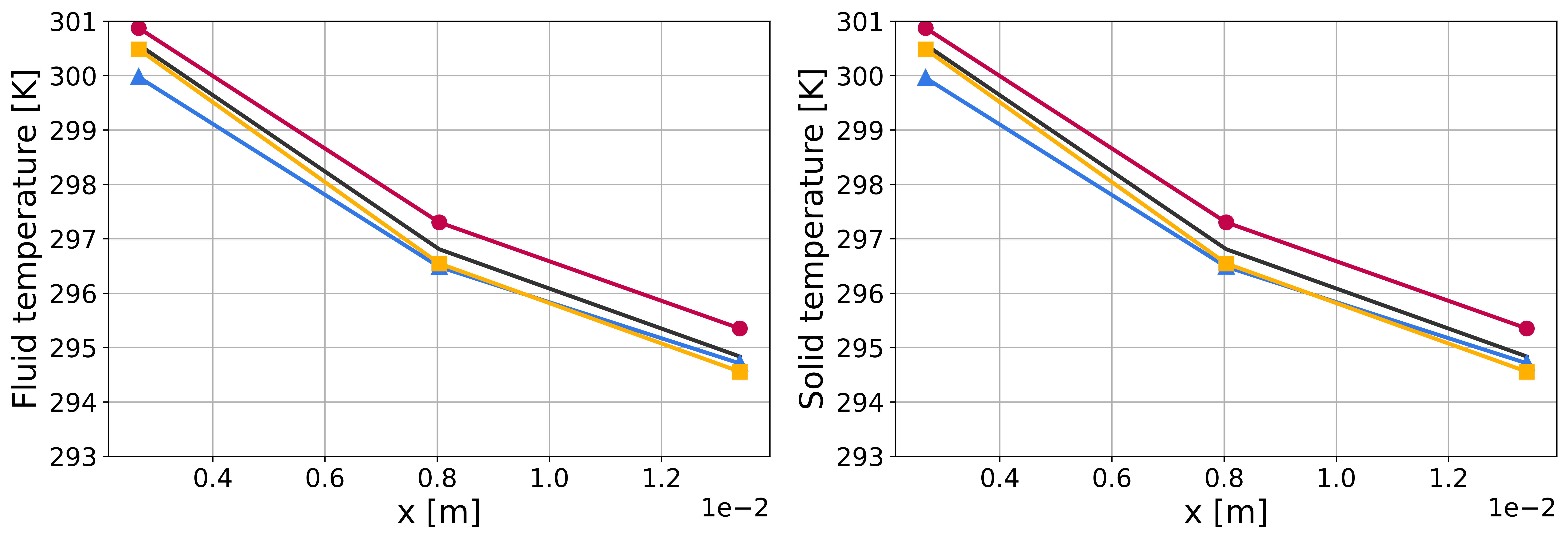}
            \caption{$t=50$s}
            \label{fig:rev_models_t2}
        \end{subfigure}
        \caption{Averaged REV-scale temperatures for three different effective conductivity models using the LTNE equations (see \Cref{sec:rev_ltne_models}). Additionally, results for the pore-resolved simulations are shown. In case of the LTNE REV-scale model with values from homogenization as well as the pore-resolved LTNE model, the heat transfer coefficient is considered as $h_1=8.3\times10^7 ~\frac{\text{W}}{\text{m}^2\text{K}}$.}
        \label{fig:rev_models}
    \end{figure}

    As stated in \Cref{sec:model_rev}, there are a variety of effective thermal conductivity models available for the REV scale. In order to consider LTNE, we presented in \Cref{sec:rev_ltne_models} formulations for the effective thermal conductivities after \citet{nuske2015modeling} and \citet{nakayama_two-energy_2001}. We obtain an additional set of effective thermal conductivities through homogenization (see \ref{app:homogenization_nonEq}). Those three different REV-scale models, will in the following be denoted by \emph{Nuske}, \emph{Nakayama} and \emph{homogenization} respectively.
    In order to compare the models, certain choices have to be made.
    As the median grain diameter $d_{50}$ is a common choice for the characteristic length $L_{ch}$ in the formulation of Nuske (\Cref{eq:rev_effLambdaI_Nuske}), the calculated diameter for the solid bodies of the dual-network model, $d_{50}\approx 3.6 \times 10^{-4}$m, is taken. For the interfacial term obtained through homogenization, the heat transfer coefficient is chosen as $h_1= 8.3\times10^7 ~\frac{\text{W}}{\text{m}^2\text{K}}$. It is important to note that the interfacial resistance is not incorporated in the formulations after Nakayama and Nuske.
    Using these three approaches to model LTNE processes on the REV scale, the results for averaged temperature profiles over space are presented in \Cref{fig:rev_models}.
    In addition to the REV-scale models, the pore-resolved results in case of $h_1$ are shown in \Cref{fig:rev_models}, as they are considered to provide the most detailed results.

    Note that, as already pointed out in \Cref{sec:case1}, the resulting solid and fluid temperatures are equal, indicating that LTE is valid.
    For an earlier time $t_1=5$s, the resulting temperatures for the different REV models differ mainly at the left boundary, where the higher fluid temperature is fixed. The difference between the models becomes slightly smaller for a later time $t_2=50$s. In this case, the difference is nearly constant over space between the formulation of Nuske and the homogenization. The approach of Nakayama leads to a slightly more pronounced temperature gradient between the evaluation points compared to the other two approaches. The discrepancies can be mainly addressed to the differing effective conductivity values resulting from the different approaches, rather than to the different interfacial heat exchange formulations. This is due to the assumption that in thermal equilibrium the heat exchange term is assumed to be negligible.
    Despite the differences between the models, all models show a similar trend in the average temperatures, which are close to the results of a pore-resolved model. The model developed by \citet{nakayama_two-energy_2001} appears to align most closely to the pore-resolve results. However, as the pore-resolved results are not fully converged in space, a slightly larger difference is expected between the effective formulation of Nakayama and the fully converged pore-resolved results (cf. \ref{app:conv_pore}).

    The REV-scale models after Nuske and Nakayama do not account for a heat transfer coefficient or an interfacial resistance. Consequently, they will not result in LTNE for cases where purely conductive systems are considered. However, this is different for homogenization, where the underlying pore-resolved equations can consider a temperature jump at the sharp fluid-solid interface depending on the heat transfer coefficient. This coefficient then appears in the effective heat exchange term on the REV scale as well.
    
\newpage
\section{Final remarks} \label{sec:finalRemarks}

    In this paper, we have addressed three research questions.
    First, comparing a dual-network and an REV-scale model with the pore-resolved reference, we showed that the REV models with the effective parameters obtained through homogenization theory are in close agreement with the pore-resolved results for both high and low interfacial resistance. Moreover, the REV-scale models indicated a similar trend in terms of the resulting temperature gradient between the evaluation points as obtained for the pore-resolved model. 
    The dual network model showed larger temperature differences and less steep temperature gradients within the system. However, the network model resulted in similar temperatures as compared to an REV model with a comparably coarse grid.
    This points out the importance of resolving the locations with large differences in phase temperatures.

    Second, regarding differences between LTE and LTNE formulations in case of different interfacial resistances, our investigation aligns with previous findings. For low Kapitza resistances at the fluid-solid interface the system remains in LTE, leading to nearly identical results for LTE and LTNE models.
    In contrast, for high interfacial resistances, corresponding to lower Biot numbers, solid and fluid temperatures differ. This is only captured by LTNE models that incorporate the interfacial resistance.

    Third, evaluating three different REV-scale LTNE models showed the deficiency of the formulations after \citet{nuske2015modeling} and \citet{nakayama_two-energy_2001}, which do not take the interfacial heat transfer coefficient into account. Hence, for purely conductive cases, those models do not result in LTNE for high interfacial resistance. In contrast, the homogenization approach includes the heat transfer coefficient and was shown to accurately capture this LTNE behavior.

    For most systems that are of interest for the investigation of LTNE, convection is present. We intend to investigate the models in the case of mass transport, and thus also heat transport due to convection, for LTNE cases in a following study.

\newpage
\appendix
\section{Homogenization from pore to REV scale} \label{app:homogenization_nonEq}
As we consider a periodic porous geometry, homogenization can be applied to derive REV-scale models from the pore-scale model equations. The pore-resolved model equations presented in \Cref{sec:model_pore} have already been homogenized by \cite{auriault2010homogenization}. We here present the main steps and resulting upscaled equations, and refer to \cite{auriault2010homogenization} for further details.

\subsection{Assumptions}
The starting point is the domain shown in \Cref{fig:setup}, where one cubic cell is repeated. As in \Cref{sec:models}, the total fluid and solid domains are denoted $\Omega_{\f}$ and $\Omega_{\s}$, and we use the model equations as stated in \Cref{sec:model_pore}. We will through the homogenization arrive at an REV-scale (upscaled) domain, where fluid and solid cannot be separated, and we denote the total, upscaled domain $\Omega$. As we will mostly refer to the fluid domain and solid domain inside one cubic reference cell $Y$, we introduce the notation $P$ and $S$ for fluid and solid inside $Y$, and $G$ for the fluid-solid interface inside $Y$.  

When performing homogenization, the underlying assumption is that one can identify a length-scale ratio $\varepsilon$ which is then allowed to approach zero. The natural choice for $\varepsilon$ is in our case $\varepsilon=\frac{l_x}{L_x}=\frac{1}{n_x}$. Letting $\varepsilon\to 0$ hence corresponds mathematically to letting gradually more (and smaller) cells fill up the domain, but in practice we are using a fixed $\varepsilon=\frac{1}{45}$. This means that the REV-scale model is only an approximation, but can still give a good approximation if $\varepsilon$ is \emph{small enough}. Earlier comparisons with heat conduction through porous media have shown good correspondence between averaged pore-scale results and REV-scale results already for $\varepsilon=0.1$ \citep{BringedalLena}. 

We further assume that we can separate the scales by introducing two coordinate systems. The macro scale (REV scale), where $L_x$ is the dominant length scale, will use $\bf x$, while the micro scale (pore scale), where $l$ is the dominant length scale, will use $\bf y$. This way, $\bf x$ will tell us where in the REV-scale domain to zoom in, while $\bf y$ is the zoomed-in coordinate resolving the detailed variability. These two coordinates systems are connected via $\varepsilon$. 

We finally assume that the model variables, in our case $T_{\f},T_{\s}$, can be expanded in case of $\varepsilon$ using two-scale asymptotic expansions. That is,
\begin{equation*}
    T_{\f}(t,{\bf x}) = T_{\f,0}(t,{\bf x},{\bf y})+\varepsilon T_{\f,1}(t,{\bf x},{\bf y}) + \varepsilon^2 T_{\f,2}(t,{\bf x},{\bf y})+\dots 
\end{equation*}
and similarly for $T_{\s}$. Note that by letting the $T_{\f,i}$ variables depend explicitly on both $\bf x,\bf y$, gradients need to be rewritten. Hence
\begin{equation*}
    \nabla T_{\f} = (\nabla_{\bf x} + \frac{1}{\varepsilon}\nabla_{\bf y})(T_{\f,0}(t,{\bf x},{\bf y})+\varepsilon T_{\f,1}(t,{\bf x},{\bf y}) + \varepsilon^2 T_{\f,2}(t,{\bf x},{\bf y})+\dots)
\end{equation*}
Then, by inserting the two-scale asymptotic expansions and identifying the dominating terms as $\varepsilon\to 0$, we can find the effective equations at the REV scale as well as effective parameters that are calculated using the reference cell. Details of the procedure is found in \citet{auriault2010homogenization}, while the resulting upscaled equations are given below.

\subsection{Homogenization of pore-scale thermal equilibrium model}\label{app:hom_equil}
We here perform the homogenization on the model from \Cref{sec:model_pore_equil}. Since $T_{\f}=T_{\s}$ on the internal boundary, the conduction will cause $T_{\f}=T_{\s}$ on every reference cube. That means that the upscaled $T_{\f}$ and $T_{\s}$ will be the same temperature $T$, and only one equation is needed to describe it. Furthermore, the dominating temperature $T_0$ will only depend on $\bf x$. The resulting REV-scale equation is therefore,
\begin{equation*}
	\partial_t(\Phi \rho_{\f}c_{\f}T + (1-\Phi)\rho_{\s} c_{\s} T) = \nabla_{\bf x}\cdot(\Lambda_\text{eff}\nabla_{\bf x} T) \quad \text{in } \Omega,
\end{equation*}
where subscript 0 has for convenience been removed. 
The matrix $\Lambda_\text{eff}$ is the effective heat conductivity, and accounts for the medium's ability to conduct heat through combined fluid and solid domains. The heat conductivity in each domain and how the phases are connected are accounted for. A full matrix is provided by the homogenization procedure, to account for any anisotropy in the geometry. Due to the chosen isotropic geometry, our $\Lambda_\text{eff}$ will in practice be scalar. The components $\Lambda_{\text{eff},i,j}$ are given by
\begin{equation*}
	\Lambda_{\text{eff},i,j} = \frac{\lambda_{\f}}{|Y|}\int_P (\delta_{ij}+\partial_{y_i}\Theta_{\f}^j({\bf y}))d{\bf y} + \frac{\lambda_{\s}}{|Y|}\int_S(\delta_{ij}+\partial_{y_i}\Theta_{\s}^j({\bf y}))d{\bf y} ,
\end{equation*}
where $\Theta_{\f}$ and $\Theta_{\s}$ are found through the cell problem
\begin{align*}
	\nabla_{\bf y}^2\Theta_{\f}^j = 0 & \quad \text{in } P, \\
	\nabla_{\bf y}^2\Theta_{\s}^j = 0 & \quad \text{in } S, \\
	\lambda_{\f}(\mathbf e_j+\nabla_{\bf y}\Theta_{\f}^j)\cdot\mathbf n = \lambda_{\s}(\mathbf e_j+\nabla_{\bf y}\Theta_{\s}^j)\cdot\mathbf n & \quad \text{on } G, \\
	\Theta_{\f}^j = \Theta_{\s}^j & \quad \text{on } G, \\
	\Theta_{\f}^j,\Theta_{\s}^j \text{ are periodic} & \quad \text{on }\partial Y.
\end{align*}
Note that the cell problem is solved on a single reference cell $Y$. 
We solve the cell problem using Netgen and NGSolve, by the same procedure as described in \Cref{sec:model_pore_solve}. For our geometry and material conductivity values, the effective matrix is
\begin{equation*}
	\Lambda_\text{eff} = \begin{pmatrix}
		2.41111 & 6.2\times 10^{-6} & 2.2\times 10^{-6}\\
		6.2\times 10^{-6} & 2.41112 & 4.5\times 10^{-6} \\
		2.2\times 10^{-6} & 4.5\times 10^{-6}& 2.41111
	\end{pmatrix} ~\frac{\mathrm{W}}{\mathrm{mK}}.
\end{equation*}
We observe that the effective heat conductivity is in practice the (scalar) value $\lambda_\text{eff}=2.4111~\frac{\mathrm{W}}{\mathrm{mK}}$, as the deviations are due to numerical errors when solving the discretized cell problems.

\subsection{Homogenization of pore-scale thermal non-equilibrium model}
We now consider the model equations from \Cref{sec:model_pore_nonequil}. 
Depending on how dominant $h$ is compared to the heat conduction, \citet{auriault2010homogenization} show that one can obtain REV-scale thermal non-equilibrium in some cases. This depends on the size of the Biot number 
\begin{equation*}
    \mathrm{Bi} = \frac{hL_x}{\lambda}.
\end{equation*}
For the values of $h, L_x$ and $\lambda$ considered in \Cref{sec:comparison}, we have that case 1 corresponds to a Biot number of $2.0\times 10^6$, and case 2 to $2.4$ when using the fluid heat conductivity. The large Biot number of case 1 leads to an REV-scale model with LTE, as in \ref{app:hom_equil}. The smaller Biot number of case 2 can place us in the regime of case IV of \cite{auriault2010homogenization}, which leads to two coupled temperature fields at the REV scale, corresponding to the model presented in Section \ref{sec:model_rev}. Then, the upscaled temperatures $T_{\f,0},T_{\s,0}$ depend only on $\bf x$ and the upscaled equations are (dropping again subscript 0 for convenience):
\begin{align*}
	\partial_t(\Phi \rho_{\f} c_{\f} T_{\f}) &= \nabla_{\bf x}\cdot(\Lambda_{\mathrm{eff,f}} \nabla_{\bf x} T_{\f}) + H(T_{\f}-T_{\s}) && \quad \text{in } \Omega,\\
	\partial_t((1-\Phi) \rho_{\s} c_{\s} T_{\s}) &= \nabla_{\bf x}\cdot(\Lambda_{\mathrm{eff,s}} \nabla_{\bf x} T_{\s}) -H(T_{\f}-T_{\s}) && \quad \text{in } \Omega.
\end{align*}
The two matrices $\Lambda_{\mathrm{eff,f}}$ and $\Lambda_{\mathrm{eff,s}}$ describe the effective heat conductivity in each of the fluid and solid domains. They account for the internal (material) heat conductivity, and also the geometry of the phase. Note that if one phase was disconnected, the effective heat conductivity would be zero. The components $\Lambda_{\mathrm{eff,f},i,j}$ and $\Lambda_{\mathrm{eff,s},i,j}$ are given by
\begin{align*}
	\Lambda_{\mathrm{eff,f},i,j} = \frac{\lambda_{\f}}{|Y|}\int_P (\delta_{ij}+\partial_{y_i}\Theta_{\f}^j({\bf y}))d{\bf y} ,\\
	\Lambda_{\mathrm{eff,s},i,j} = \frac{\lambda_{\s}}{|Y|}\int_S (\delta_{ij}+\partial_{y_i}\Theta_{\s}^j({\bf y}))d{\bf y},
\end{align*}
where $\Theta_{\f}$ and $\Theta_{\s}$ are solved through two (uncoupled) cell problems
\begin{align*}
\nabla_{\bf y}^2\Theta_{\f}^j = 0 & \quad \text{in } P, \\
(\mathbf e_j+\nabla_{\bf y}\Theta_{\f}^j)\cdot\mathbf n = 0 &\quad \text{on } G,\\
\Theta_{\f}^j \text{ is periodic} &\quad \text{on }\partial Y,
\end{align*}
and
\begin{align*}
	\nabla_{\bf y}^2\Theta_{\s}^j = 0 & \quad \text{in } S, \\
	(\mathbf e_j+\nabla_{\bf y}\Theta_{\s}^j)\cdot\mathbf n = 0 &\quad \text{on } G, \\
	\Theta_{\s}^j \text{ is periodic} &\quad \text{on }\partial Y.
\end{align*}
For our geometry and material conductivity values, the effective matrices are
\begin{equation*}
	\Lambda_{\mathrm{eff,f}} = \lambda_{\f}\begin{pmatrix}
		0.108767 & 9.8\times 10^{-8} & 4.6\times 10^{-7}\\
		9.8\times 10^{-8} & 0.108771 & 1.3\times 10^{-6} \\
		4.6\times 10^{-7} & 1.3\times 10^{-6}& 0.108767
	\end{pmatrix},
\end{equation*}
\begin{equation*}
	\Lambda_{\mathrm{eff,s}} = \lambda_{\s}\begin{pmatrix}
		0.863663 & -1.9\times 10^{-7} & -1.1\times 10^{-7}\\
		-1.9\times 10^{-7} & 0.863661 & 1.8\times 10^{-7} \\
		-1.1\times 10^{-7} & 1.8\times 10^{-7}& 0.863662
	\end{pmatrix}.
\end{equation*}
Again we obtain in practice scalar values $\lambda_{\mathrm{eff,f}}=0.0739~\frac{\mathrm{W}}{\mathrm{mK}}$ and $\lambda_{\mathrm{eff,s}}=2.418~\frac{\mathrm{W}}{\mathrm{mK}}$ for the effective heat conductivities of the two phases.

The effective parameter $H$, that couples the upscaled $T_{\f}$ and $T_{\s}$, is given by
\begin{equation*}
	H=\frac{1}{|Y|}\int_G hd{\bf y} = \frac{h}{|Y|}\int_G d{\bf y},
\end{equation*}
that is, it is the pore scale $h$ multiplied with specific surface area, which is denoted by $a_{\f\s}$ at the REV scale. 

\newpage
\section{Additional information for the constructed dual-network model} \label{app:dnm_additional}
\subsection{Choice of shape parameters} \label{sec:DNM_shape_parameters}
    The shape parameters for conductive energy exchange, $C_{0,\f},C_{0,\s}, C_{\infty,\f}, C_{\infty,\s}$ and $C_{\mathrm{I}}$, can be obtained through comparing  effective thermal phase conductivities for the dual-network model to the ones of the pore-resolved simulations. For this, a temperature gradient $\Delta T$ for the solid and the fluid phase is applied in the direction of interest (for an isotropic, homogeneous medium, one direction is enough to consider), while all other boundary conditions are set to zero-Neumann. Evaluating the heat flux $q_{\mathrm{outflow}}$ from the fluid and the solid phase, the effective thermal conductivities can be obtained through
    \begin{equation}
        \lambda_{\mathrm{eff}, \f} = \frac{q_{\mathrm{outflow},\f}}{A_{\mathrm{outflow}}\Delta T} \, , \quad \lambda_{\mathrm{eff}, \s} = \frac{q_{\mathrm{outflow},\s}}{A_{\mathrm{outflow}}\Delta T} \; .
    \end{equation}
    These values can be consequently compared to those obtained through an analogous pore-scale simulation or to values obtained through homogenization. Given a definite number of shape parameter sets, the set with the minimal summed relative error can be chosen as the best fit. The following intervals were chosen for the respective shape parameters:
    \begin{align*}
        &C_{0,\f} \in (0,1] \, , \, C_{\infty,\f} = \max \left(1, f_{C\infty,\f}\nicefrac{A_{i,\f}}{A_{ij,\f}}\right) \text{ with } f_{C\infty,\f} \in \{0\}\cup \left(\nicefrac{A_{ij,\f}}{A_{i,\f}},1\right]\, , \\
        &C_{0,\s} \in (0,1] \, , \, C_{\infty,\s} = \max \left(1, f_{C\infty,\s}\nicefrac{A_{i,\s}}{A_{ij,\s}}\right) \text{ with } f_{C\infty,\s} \in \{0\}\cup \left(\nicefrac{A_{ij,\s}}{A_{i,\s}},2.5\right]\, , \\
        &C_I \in (0,1) \; .
    \end{align*}
    For the comparison study in \Cref{sec:comparison}, the best fit for the shape parameters was obtained as
    \begin{align*}
        C_{0,\f} &= 0.4\, , &&\quad C_{\infty,\f} = \max \left(1, 0.2\times\nicefrac{A_{i,\f}}{A_{ij,\f}}\right) \, ,\\ C_{0,\s} &= 0.05 \, , &&\quad C_{\infty,\s} = \max \left(1, 2.25 \times \nicefrac{A_{i,\s}}{A_{ij,\s}}\right)  \, ,\\
        C_{\mathrm{I}} &= 0.4\;.
    \end{align*}
    Hereby, $A_{i,\s}$ is being estimated through $A_{i,\s} = \nicefrac{V_{i,\s}}{\Delta x}$,
    with the volume of the solid body $V_{i,\s}$ and the distance from the solid body center to the center of the solid connection $\Delta x$ (see \cite{koch_dual_2021}).
    The corresponding effective thermal conductivities for the bulk phases are
    \begin{equation*}
        \lambda_{\mathrm{eff},\f}^{\mathrm{DNM}} = 0.1087 \lambda_{\f}= 0.0738 ~\nicefrac{\mathrm{W}}{\mathrm{mK}}\, , \; \lambda_{\mathrm{eff},\s}^{\mathrm{DNM}} = 0.8622  \lambda_{\s}=2.4142~\nicefrac{\mathrm{W}}{\mathrm{mK}}\;. 
    \end{equation*}

\subsection{Setting boundary conditions for dual-network model} \label{sec:DNM_boundary}
\begin{figure}[h!]
        \centering
        \includegraphics[width=\linewidth]{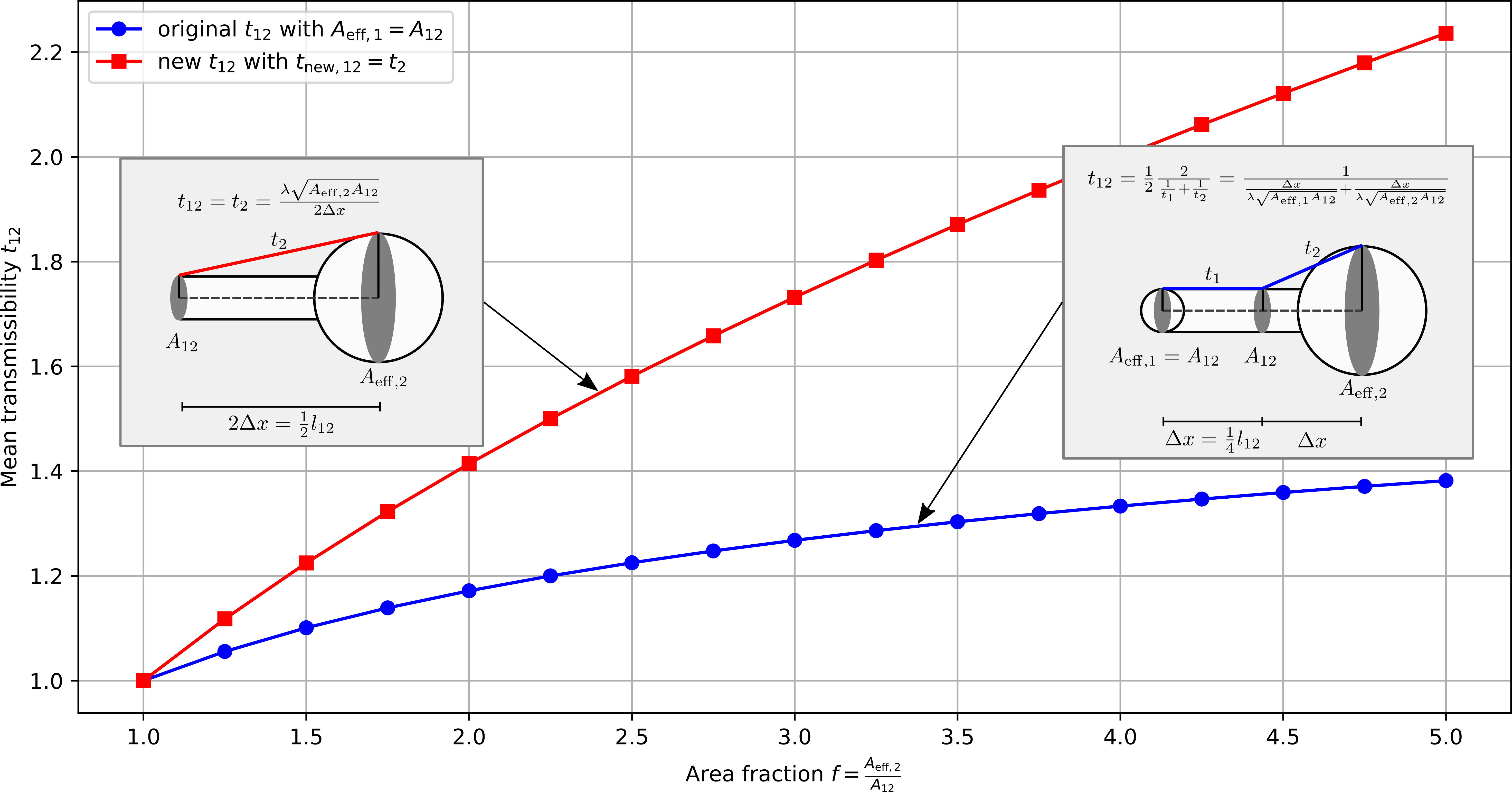}
        \caption{Different mean boundary transmissibilities for zero-volume boundary pore bodies. The effect of the different approaches for varying cross-sectional of pore body $2$ is shown.}
        \label{fig:DNM_boundary_t}
    \end{figure}

    The pore-scale geometry of the domain mentioned in \Cref{sec:comparison} leads to solid bodies directly at the domain boundaries, while the most outer pore bodies are located in the center of each reference cell facing the boundary. For the solid boundary bodies, the volume is adjusted so that bodies at the corners, edges, and faces of the surrounding box have one eighth, one fourth, and one half of the volume compared to the interior solid bodies. In \DuMuX, the boundary conditions are always set for the pore or solid bodies of a network. However, for the void grid, only pore throats, but no pore bodies, are located directly at the domain boundaries. Hence, we add void bodies at those boundaries, each connected only through the boundary throats, with half the length as the interior throats, to one interior void body. The void boundary bodies are assigned zero volume in order to neglect additional storage effects. Boundary conditions are then set directly at the boundary, rather than half a reference cell length inside the domain. Note that zero volume does not mean that the effective area of this pore body, which is used within the transmissibility formulation, is also $0$. This will result in the same area as the adjacent throat cross-sectional area, $A_{\mathrm{eff, boundary}}=A_{ij}$ (cf. \cite{koch_dual_2021}). However, using the same averaging for the thermal transmissibility as for the interior void bodies, namely a harmonic average between two cone approximations from the pore bodies to the throat centers, would lead to an underestimation of the mean thermal conductivity with respect to the interior domain. Therefore, the thermal conductivity of the boundary throats adjacent to the added boundary pore bodies is adapted to be consistent with the formulation in the interior of the domain. For this, only one cone approximation is taken instead of the harmonic mean between two cone approximations, as without added boundary pores only a throat would be attached to the boundary (see \Cref{fig:DNM_boundary_t}).

\newpage
\section{Spatial and temporal convergence of different models} \label{app:convergence}
In the following the choice of the time and space discretization in \Cref{sec:comparison} is investigated for each of the three model classes. Note that the finest discretization is shown in all figures as a solid black line.

\subsection{Pore-resolved model}\label{app:conv_pore}
\begin{figure}
        \begin{subfigure}{\linewidth}
            \centering
            \includegraphics[width=\linewidth]{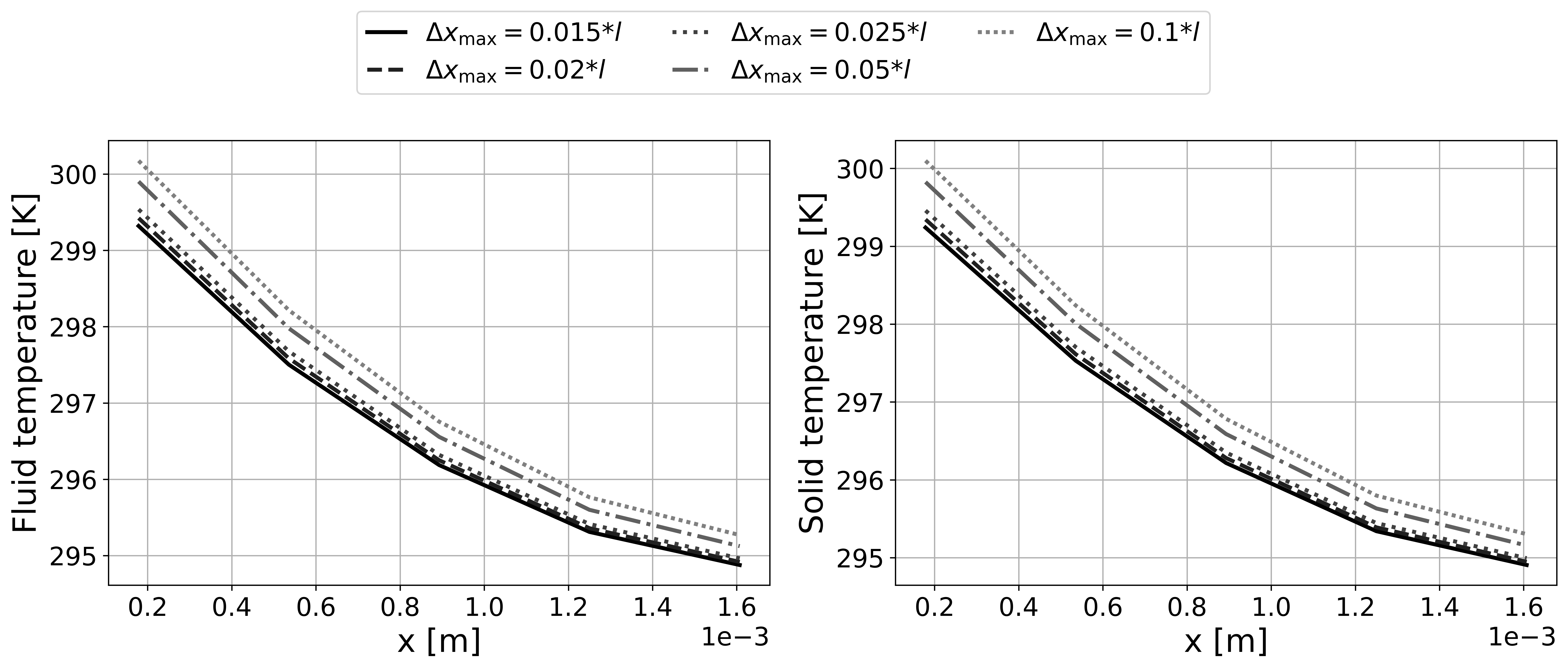}
            \caption{$t_1=1$s}
            \label{fig:pr_conv_space_t1}
        \end{subfigure}
        \begin{subfigure}{\linewidth}
            \centering
            \includegraphics[width=\linewidth]{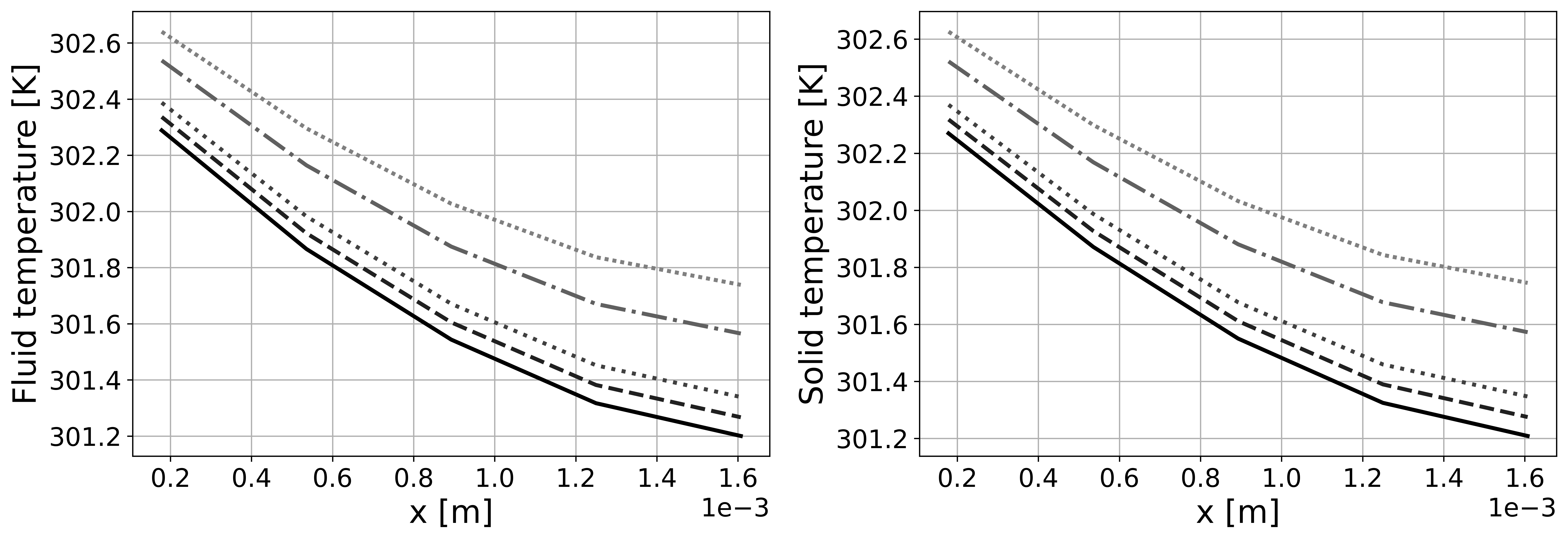}
            \caption{$t_2=5$s}
            \label{fig:pr_conv_space_t2}
        \end{subfigure}
        \caption{Temperature averaged over each of the five reference cells in horizontal direction for different maximal element sizes $\Delta x_{\mathrm{max}}$. Left is fluid temperature, right solid. Top row is for an early time, and the bottom row for a later time when the equilibrium temperature is close to be approached. Note that $l=\nicefrac{1}{2800}~\mathrm{m}$ denotes the length of a reference cell.}
        \label{fig:pr_conv_space}
    \end{figure}

\begin{figure}
        \begin{subfigure}{\linewidth}
            \centering
            \includegraphics[width=\linewidth]{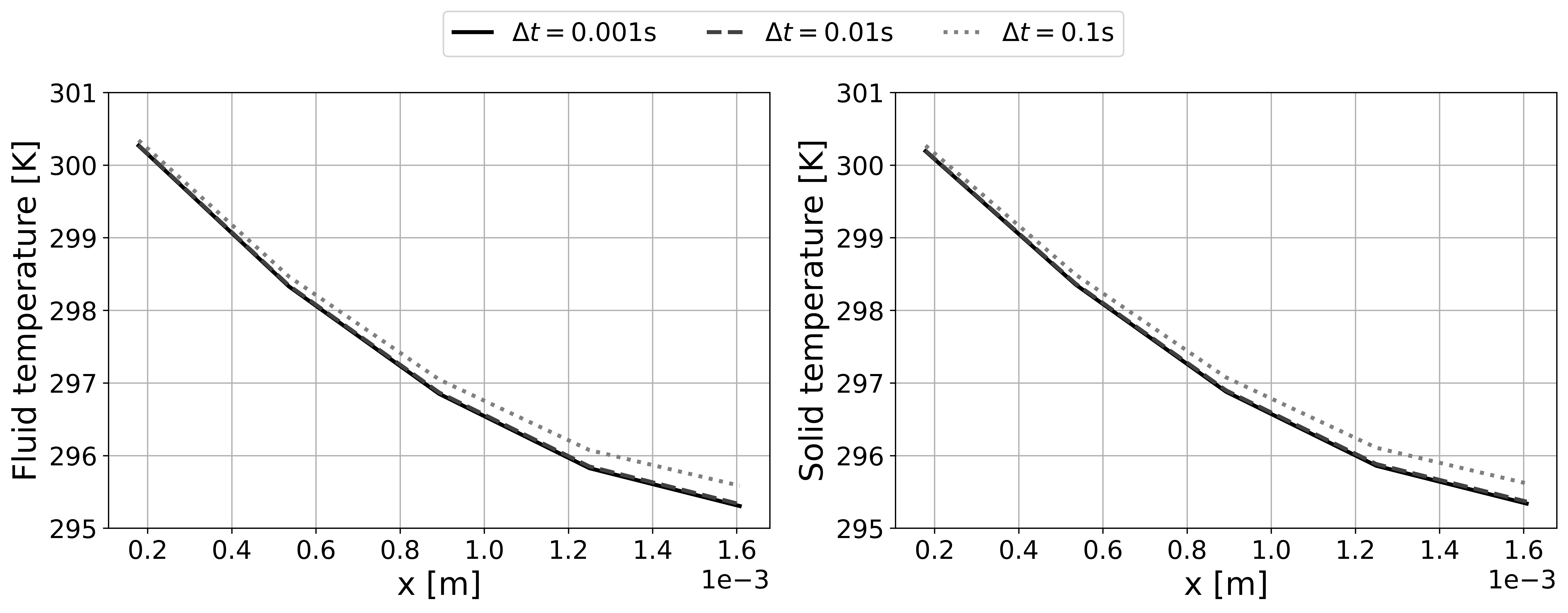}
            \caption{$t_1=1$s}
            \label{fig:pr_conv_time_t1}
        \end{subfigure}
        \begin{subfigure}{\linewidth}
            \centering
            \includegraphics[width=\linewidth]{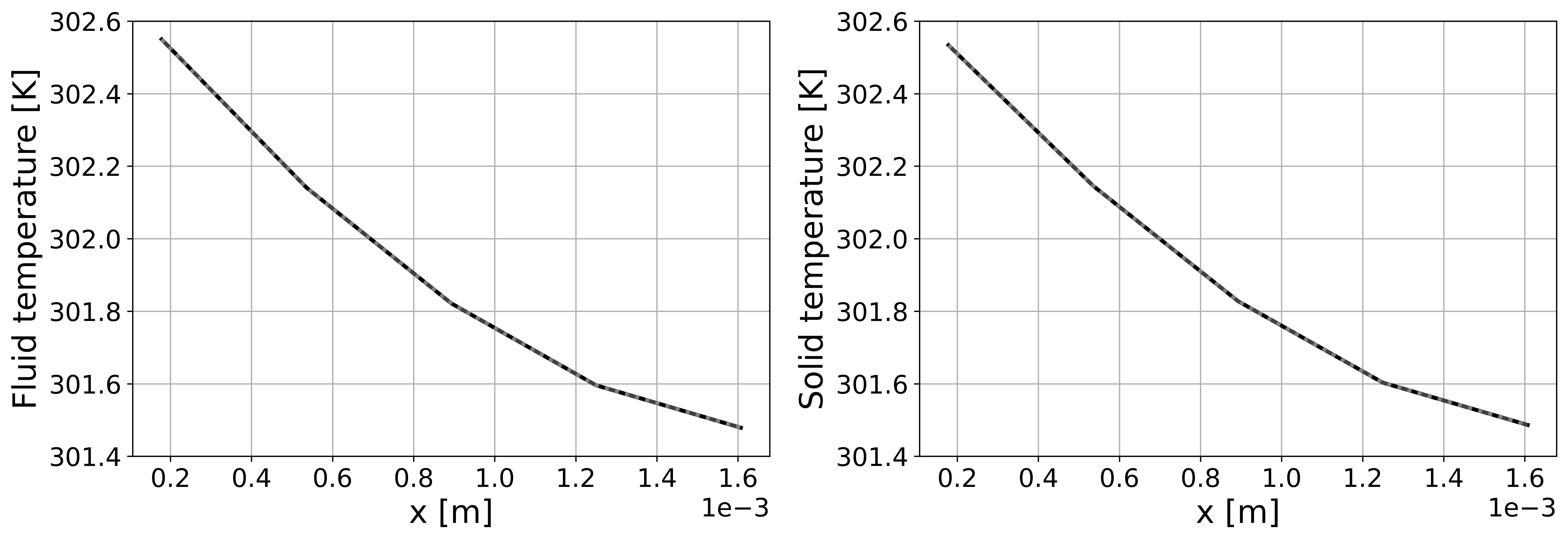}
            \caption{$t_2=5$s}
            \label{fig:pr_conv_time_t2}
        \end{subfigure}
        \caption{Temperature averaged over each of the five reference cells in horizontal direction for different time-step sizes $\Delta t$. Left is fluid temperature, right solid. Top row is for an early time, and the bottom row for a later time when the equilibrium temperature is close to be approached.}
        \label{fig:pr_conv_time}
    \end{figure}

For the convergence study of the pore-resolved model we consider a smaller domain to be able to resolve the geometry with finer meshes. We use a setup with $5\times 1\times 1$ cells, and compare the average temperatures of each cell as the mesh is refined. Additionally, we use the model equations from \Cref{sec:model_pore_equil} with temperature continuity for the convergence study. As seen from \Cref{fig:pr_conv_space}, the average temperatures change slightly as the grid is refined. The changes show that the system exhibits numerical diffusion that influences the results for the coarser meshes. The simulations in \Cref{sec:comparison} are however done with the coarsest mesh, $\Delta x_\text{max}=0.1/2800~\mathrm{m}$, due to the larger domain size. The convergence study shows that the pore-resolved results contain numerical diffusion giving somewhat overestimated conductive heat transport through the domain. The same setup with the coarsest grid is used for convergence in time. \Cref{fig:pr_conv_time} shows that the time discretization error to a small extent influences the results, for the time-step sizes considered here. In \Cref{sec:comparison}, a time-step size of $\Delta t=10^{-1}~\mathrm{s}$ has been used for all pore-resolved simulations.

\subsection{Dual-Network Model} \label{sec:conv_time_dnm}
\begin{figure}
        \begin{subfigure}{\linewidth}
            \centering
            \includegraphics[width=\linewidth]{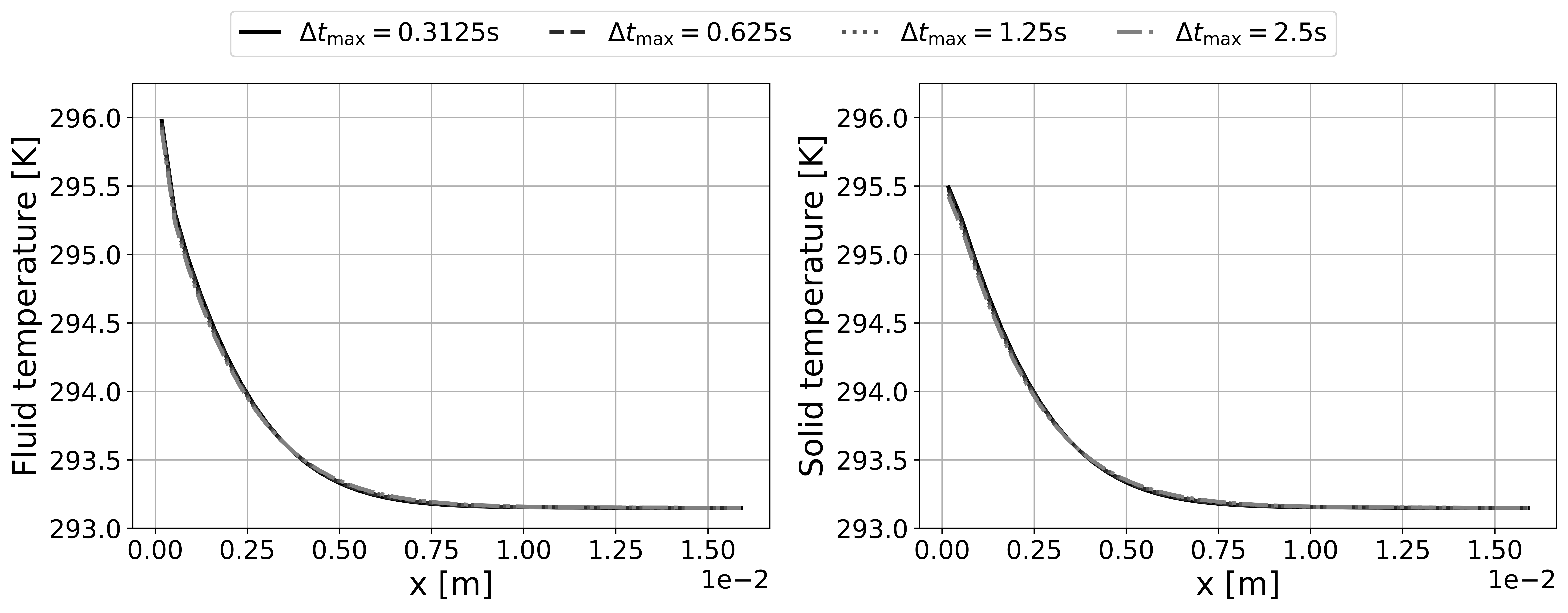}
            \caption{$t_1=5$s}
            \label{fig:dnm_conv_time_lte_t1}
        \end{subfigure}
        \begin{subfigure}{\linewidth}
            \centering
            \includegraphics[width=\linewidth]{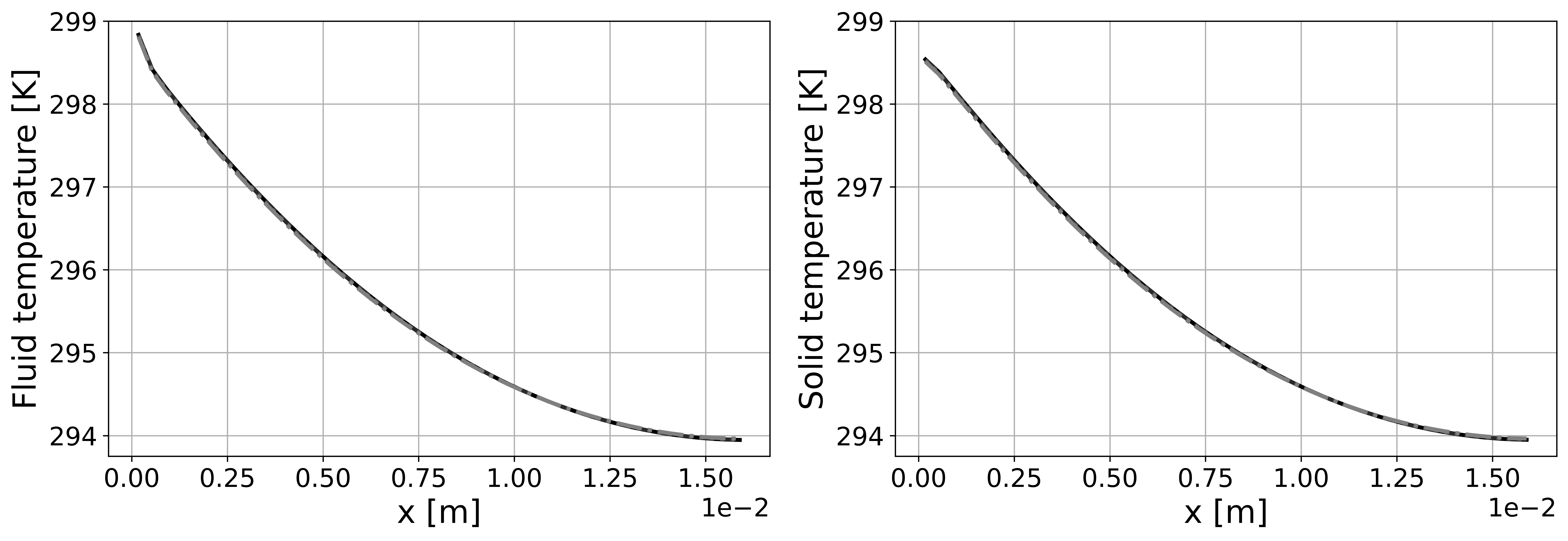}
            \caption{$t_2=50$s}
            \label{fig:dnm_conv_time_lte_t2}
        \end{subfigure}
        \caption{Average temperatures are shown resulting from the dual-network model in case of temperature continuity at the underlying sharp interface between the solid and the fluid phase for different maximum time-steps $\Delta t_{\mathrm{max}}$. Fluid and solid temperatures, on the left- and the right-hand side, are presented for one earlier time $t_1=5~\mathrm{s}$ at the top and one later time $t_2=50~\mathrm{s}$ at the bottom.}
        \label{fig:dnm_conv_time_lte}
    \end{figure}

\begin{figure}
        \begin{subfigure}{\linewidth}
            \centering
            \includegraphics[width=\linewidth]{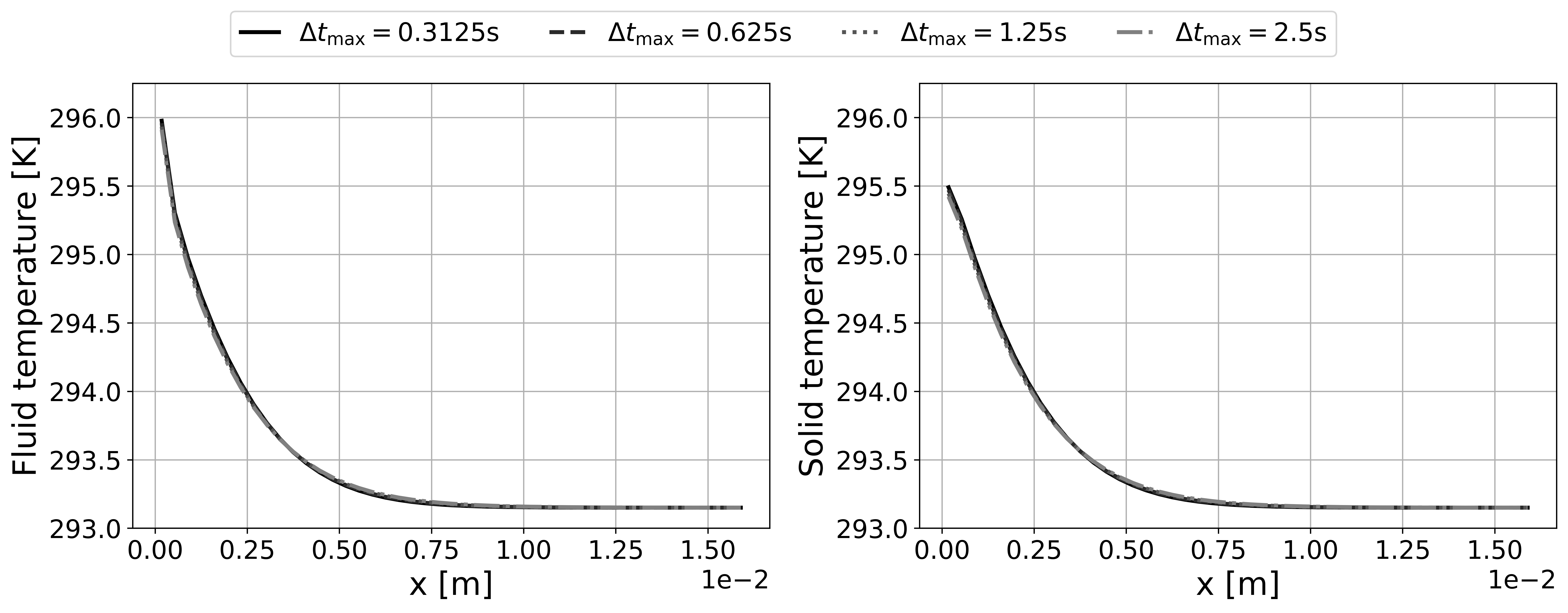}
            \caption{$t=5$s}
            \label{fig:dnm_conv_time_ltne1_t1}
        \end{subfigure}
        \begin{subfigure}{\linewidth}
            \centering
            \includegraphics[width=\linewidth]{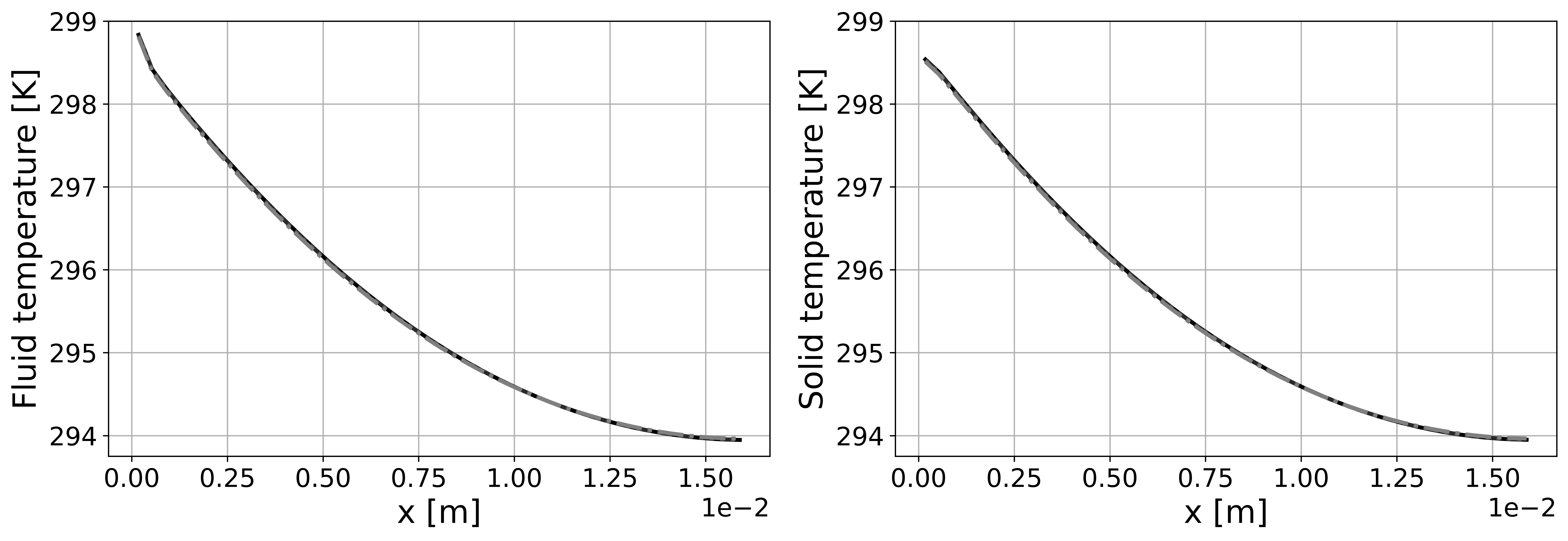}
            \caption{$t_2=50$s}
            \label{fig:dnm_conv_time_ltne1_t2}
        \end{subfigure}
        \caption{Average temperatures are shown resulting from the dual-network model in case of a heat transfer coefficient of $h=3.8\times10^7~\frac{\mathrm{W}}{\mathrm{m}^2\mathrm{K}}$ (as in \Cref{sec:case1}) for different maximum time-steps $\Delta t_{\mathrm{max}}$. Fluid and solid temperatures, on the left- and the right-hand side, are presented for one earlier time $t_1=5~\mathrm{s}$ at the top and one later time $t_2=50~\mathrm{s}$ at the bottom.}
        \label{fig:dnm_conv_time_ltne1}
    \end{figure}

\begin{figure}
        \begin{subfigure}{\linewidth}
            \centering
            \includegraphics[width=\linewidth]{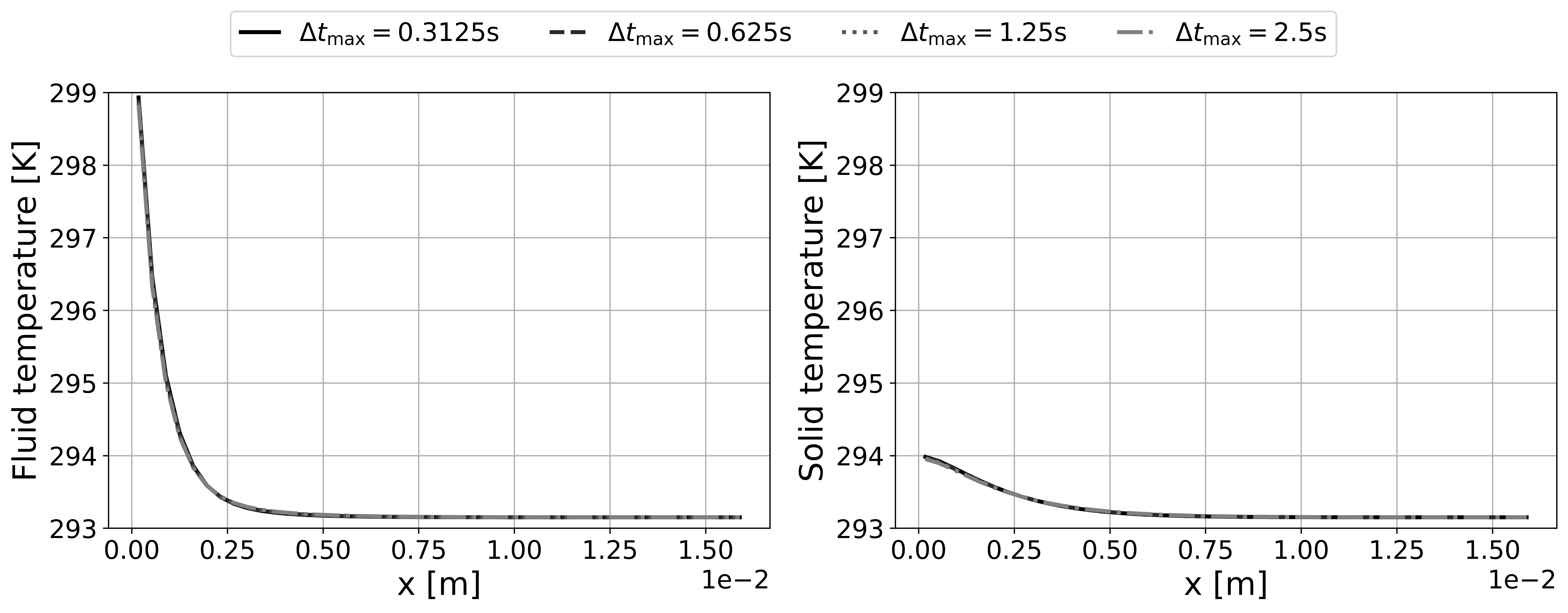}
            \caption{$t_1=5$s}
            \label{fig:dnm_conv_time_ltne2_t1}
        \end{subfigure}
        \begin{subfigure}{\linewidth}
            \centering
            \includegraphics[width=\linewidth]{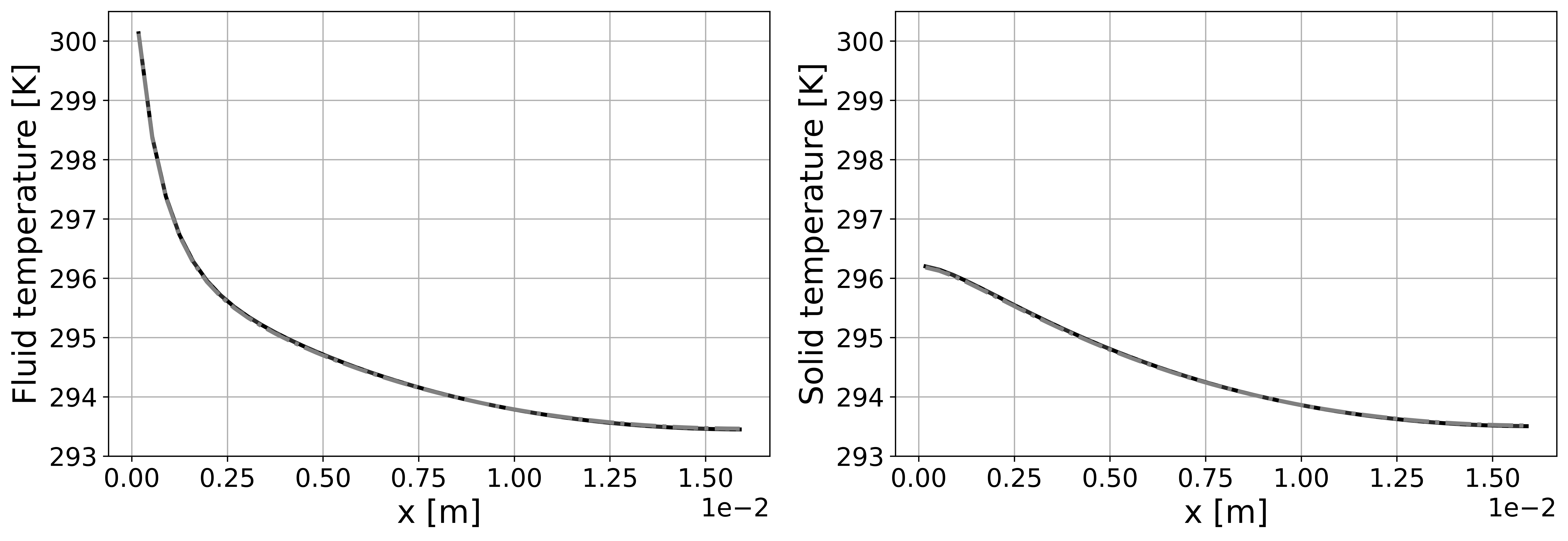}
            \caption{$t_2=50$s}
            \label{fig:dnm_conv_time_ltne2_t2}
        \end{subfigure}
        \caption{Average temperatures are shown resulting from the dual-network model in case of a heat transfer coefficient of $h=100~\frac{\mathrm{W}}{\mathrm{m}^2\mathrm{K}}$ (as in \Cref{sec:case2}) for different maximum time-steps $\Delta t_{\mathrm{max}}$. Fluid and solid temperatures, on the left- and the right-hand side, are presented for one earlier time $t_1=5~\mathrm{s}$ at the top and one later time $t_2=50~\mathrm{s}$ at the bottom.}
        \label{fig:dnm_conv_time_ltne2}
    \end{figure}

    As the discretization in space is fixed to the pore-scale geometry through the location of the void and solid bodies and hence can not be refined for pore network models, only the discretization in time will be investigated. We investigate in figures \ref{fig:dnm_conv_time_lte}, \ref{fig:dnm_conv_time_ltne1} and \ref{fig:dnm_conv_time_ltne2} the influence of the maximum time-step size $\Delta t_{\mathrm{max}}$ on the resulting temperature profiles in case of $45\times4\times4$ reference cell as in \Cref{sec:comparison}. This investigation is shown in the case of temperature continuity at the underlying sharp interface (see \Cref{fig:dnm_conv_time_lte}), with a heat transfer of $h=8.3\times 10^7 ~\frac{\mathrm{W}}{\mathrm{m}^2\mathrm{K}}$ like in \Cref{sec:case1} (see \Cref{fig:dnm_conv_time_ltne1}) and with $h=100~\frac{\mathrm{W}}{\mathrm{m}^2\mathrm{K}}$ (see \Cref{fig:dnm_conv_time_ltne2}).
    Compared to \Cref{sec:comparison}, we show here the average temperatures for each slice of reference cells in $y-$ and $z-$ direction, leading to averages over $1\times4\times4$ reference cells.
    For all three cases, there are only minimal differences between the results of the coarsest and finest maximum time-step size $\Delta t_{\mathrm{max}}$, which is due to the adaptive time-stepping scheme that is used. Therefore, time-step sizes get already refined for times, when it is needed. Consequently, a maximum time-step size of $\Delta t_{\mathrm{max}}=2.5~\mathrm{s}$ is chosen for the comparisons in \Cref{sec:comparison}.

\subsection{REV-scale model}
    For the REV-scale model, the temporal and spatial discretization is investigated hereafter. The same setup as defined in \Cref{sec:comparison} is used. Mean temperatures are evaluated for three distinct REV-cells, which are considered to contain $15\times 4 \times 4$ reference cells.
    When refining the time-step size, the number of cells in $x-$direction is fixed to the finest spatial discretization of $n_{\mathrm{cells},x} = 1440$. For the investigation of the spatial discretization, the finest time-step size $\Delta t = 0.3125 ~ \mathrm{s}$ is used accordingly. For the investigation of both spatial and time discretization, three cases are investigated. First, the results in case of equilibrium conditions, secondly in case of a heat transfer coefficient with $h=8.3\times 10^7 ~\frac{\mathrm{W}}{\mathrm{m}^2\mathrm{K}}$ and lastly with $h=100 ~\frac{\mathrm{W}}{\mathrm{m}^2\mathrm{K}}$. Note that this corresponds exactly to the same cases as investigated in \Cref{sec:comparison}.

\begin{figure}
        \begin{subfigure}{0.49\linewidth}
            \centering
            \includegraphics[width=\linewidth]{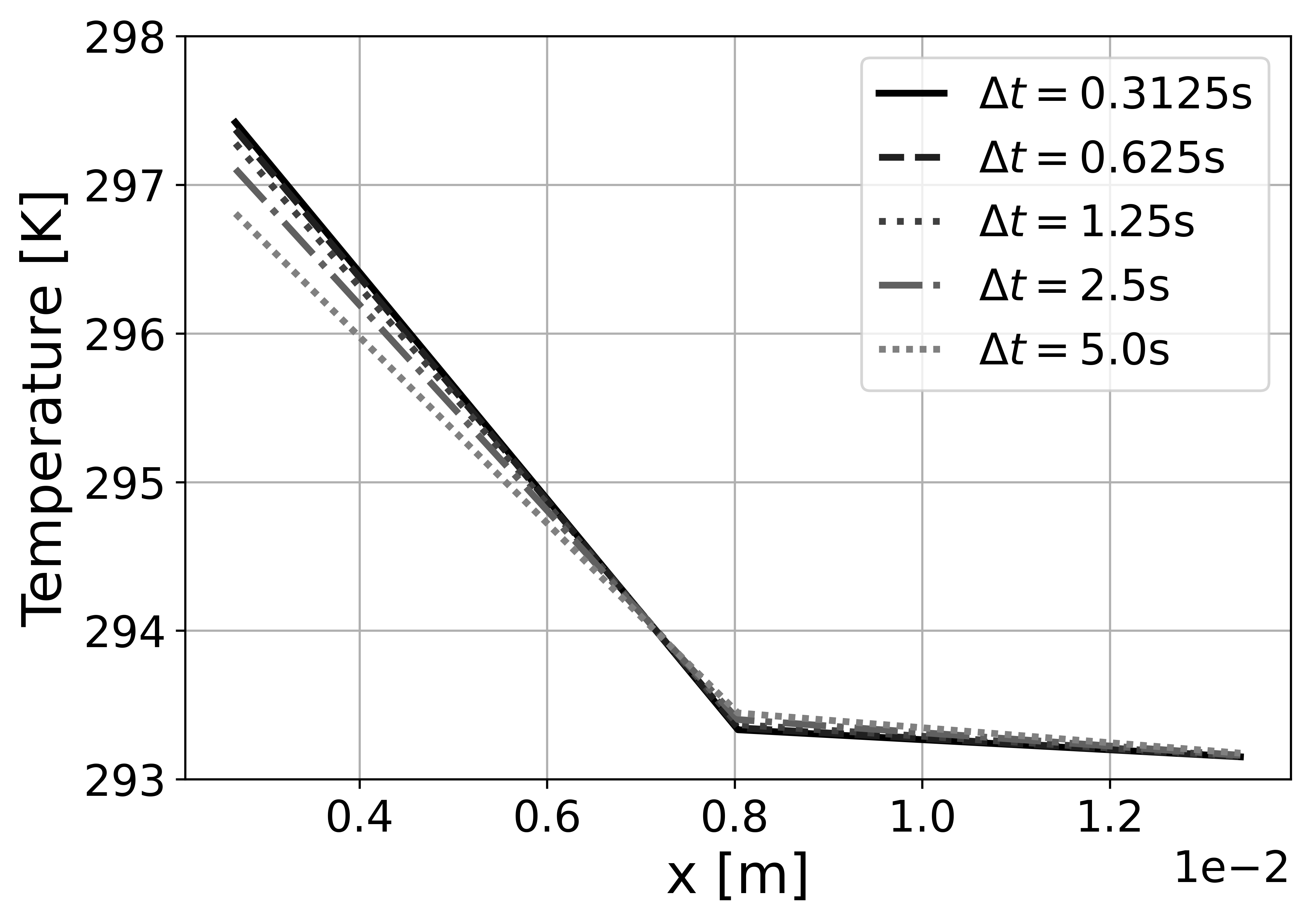}
            \caption{$t_1=5$s}
            \label{fig:rev_conv_time_lte_t1}
        \end{subfigure} \hfill
        \begin{subfigure}{0.49\linewidth}
            \centering
            \includegraphics[width=\linewidth]{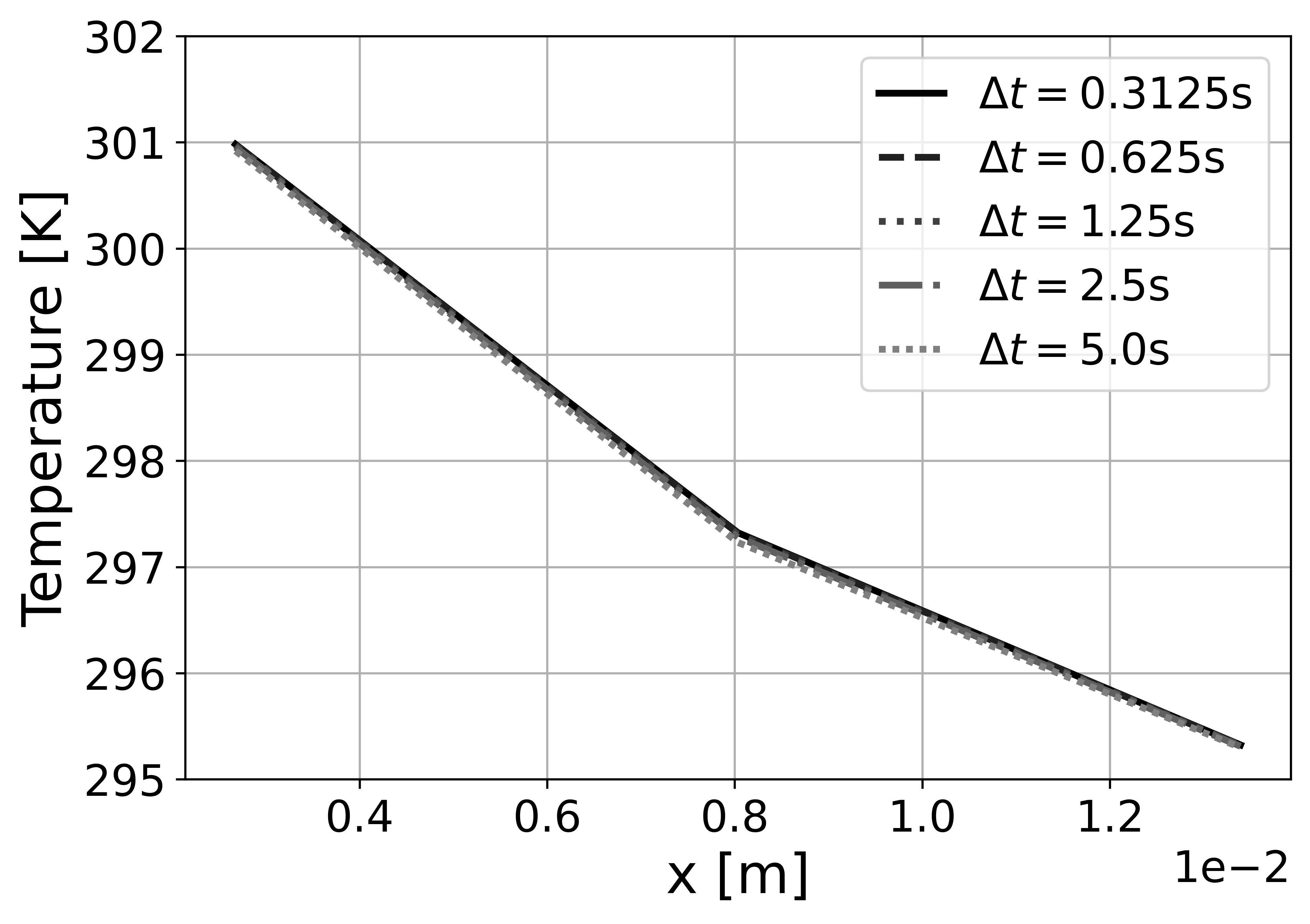}
            \caption{$t_2=50$s}
            \label{fig:rev_conv_time_lte_t2}
        \end{subfigure}
        \caption{Average temperatures for the REV-scale model in case of LTE for different fixed time-step sizes $\Delta t$. On the left one earlier time at $t_1=5~\mathrm{s}$ is shown and on the right a later time $t_2=50~\mathrm{s}$. The number of discretization cells in $x-$direction is $1440$.}
        \label{fig:rev_conv_time_lte}
    \end{figure}

\begin{figure}
        \begin{subfigure}{\linewidth}
            \centering
            \includegraphics[width=\linewidth]{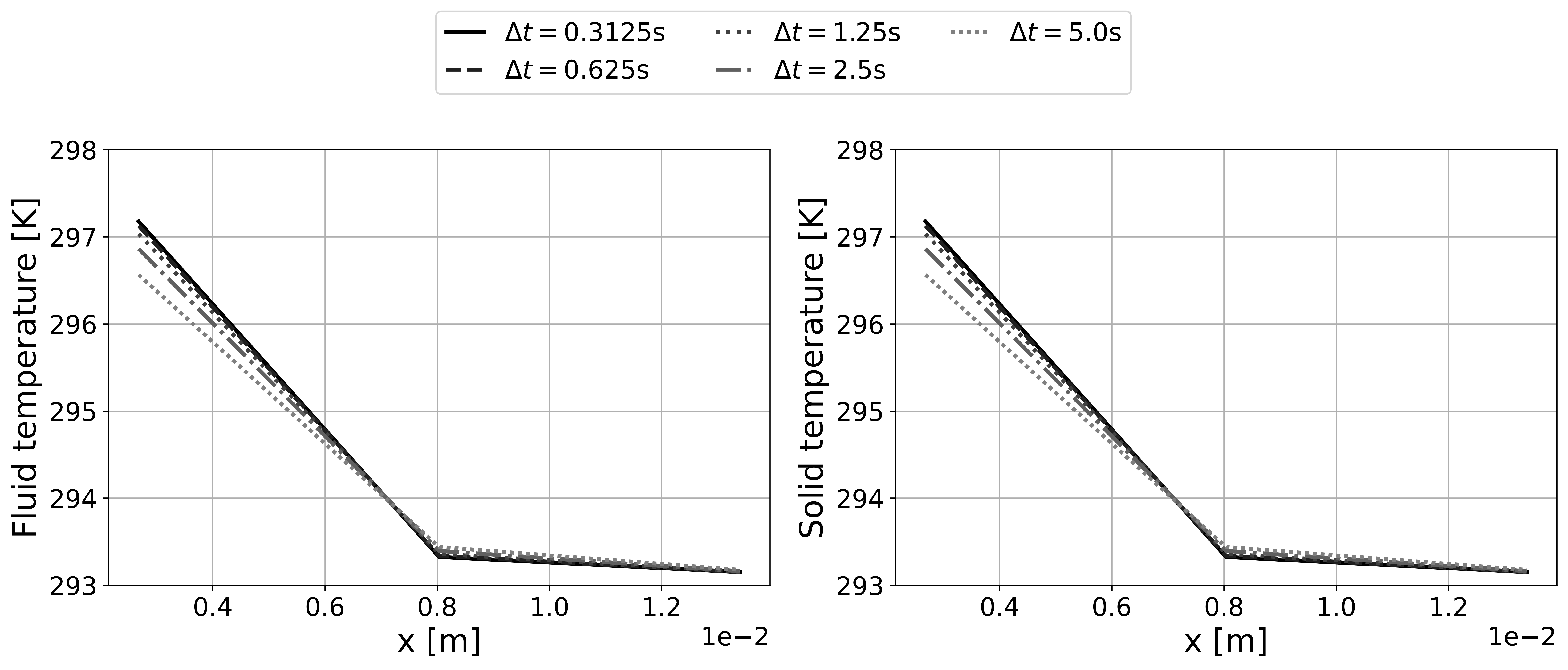}
            \caption{$t_1=5$s}
            \label{fig:rev_conv_time_ltne1_t1}
        \end{subfigure}
        \begin{subfigure}{\linewidth}
            \centering
            \includegraphics[width=\linewidth]{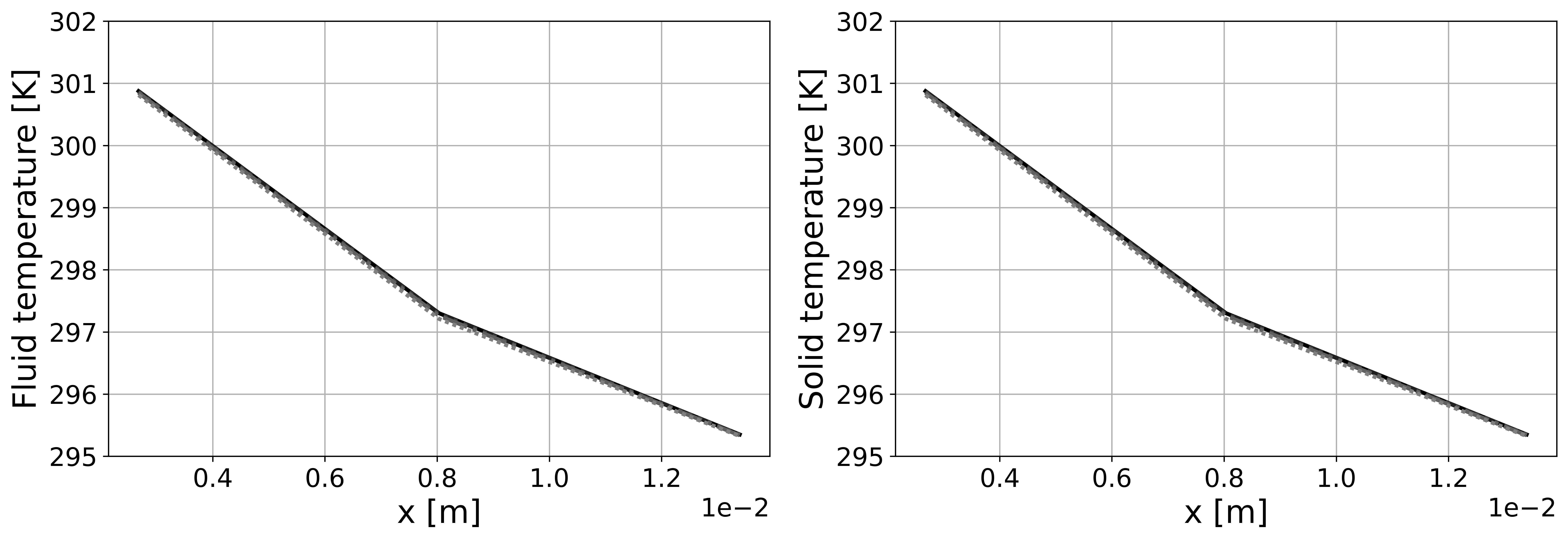}
            \caption{$t_2=50$s}
            \label{fig:rev_conv_time_ltne1_t2}
        \end{subfigure}
        \caption{Average temperatures for the REV-scale model in case of LTNE with $h=3.8\times10^7~\frac{\mathrm{W}}{\mathrm{m}^2\mathrm{K}}$ (as in \Cref{sec:case1}) for different fixed time-step sizes $\Delta t$. Fluid and solid temperatures are shown on the left and right, while one earlier time with $t_1=5~\mathrm{s}$ is shown at the top row and one later time $t_2=50~\mathrm{s}$ at the bottom row. The number of discretization cells in $x-$direction is $1440$.}
        \label{fig:rev_conv_time_ltne1}
    \end{figure}

\begin{figure}
        \begin{subfigure}{\linewidth}
            \centering
            \includegraphics[width=\linewidth]{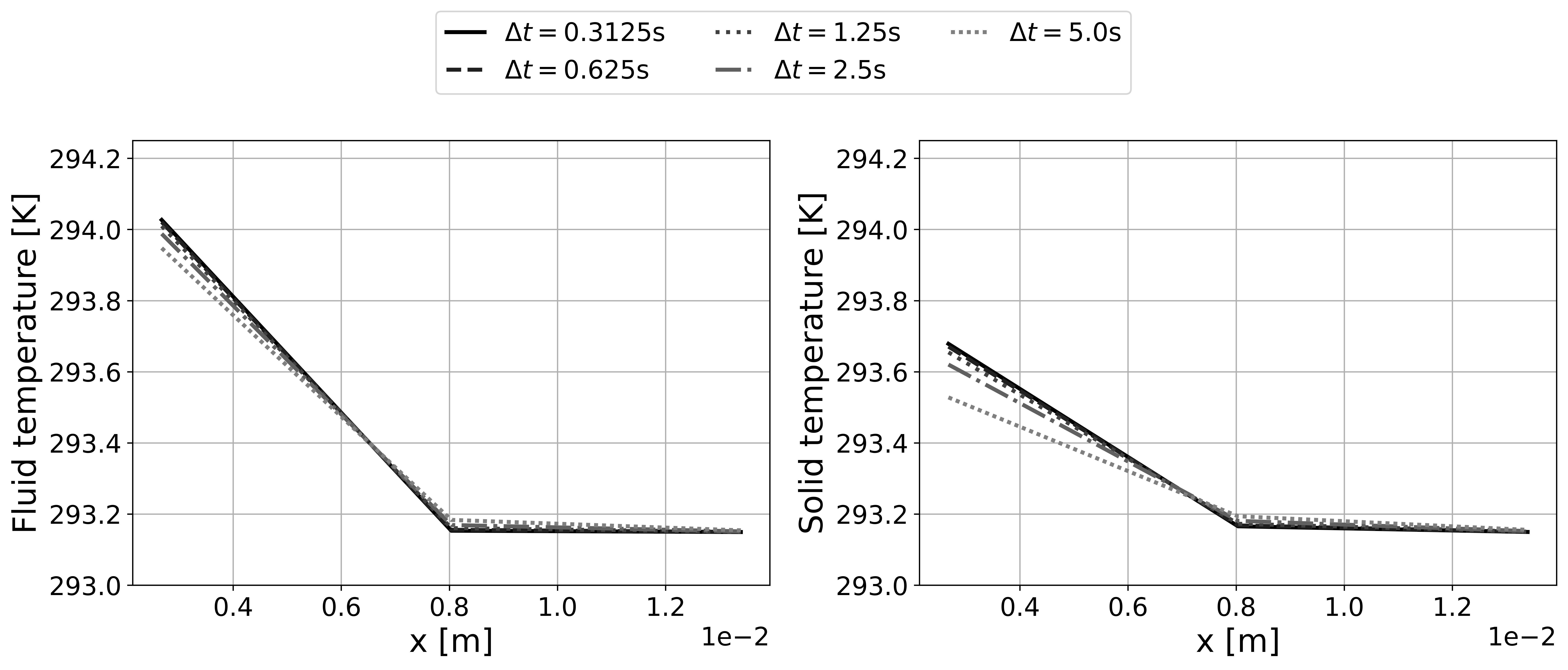}
            \caption{$t_1=5$s}
            \label{fig:rev_conv_time_ltne2_t1}
        \end{subfigure}
        \begin{subfigure}{\linewidth}
            \centering
            \includegraphics[width=\linewidth]{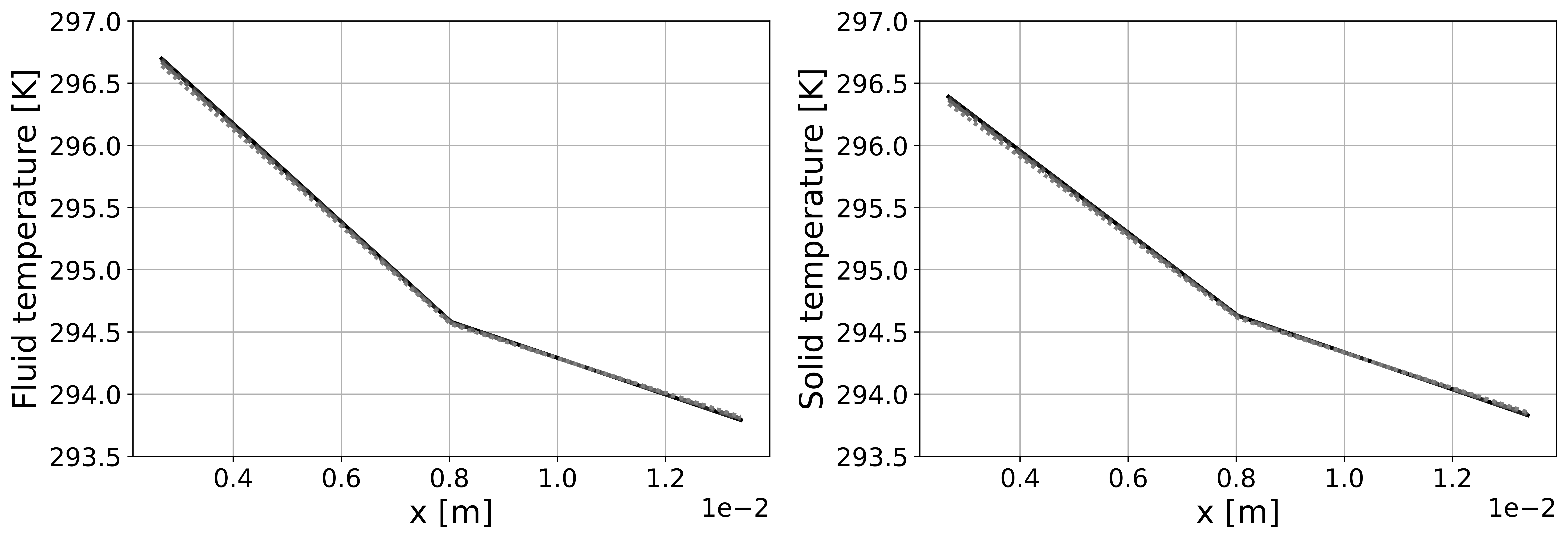}
            \caption{$t_2=50$s}
            \label{fig:rev_conv_time_ltne2_t2}
        \end{subfigure}
        \caption{Average temperatures for the REV-scale model in case of LTNE with $h=100~\frac{\mathrm{W}}{\mathrm{m}^2\mathrm{K}}$ (as in \Cref{sec:case1}) for different fixed time-step sizes $\Delta t$. Fluid and solid temperatures are shown on the left and right, while one earlier time with $t_1=5~\mathrm{s}$ is shown at the top row and one later time $t_2=50~\mathrm{s}$ at the bottom row. The number of discretization cells in $x-$direction is $1440$.}
        \label{fig:rev_conv_time_ltne2}
    \end{figure}

    Figures \ref{fig:rev_conv_time_lte}, \ref{fig:rev_conv_time_ltne1} and \ref{fig:rev_conv_time_ltne2} show the influence of refined time-steps on the resulting REV-scale temperatures for time $t_1=5~$s and $t_2=50~$s. Although for a smaller time $t_1=5~\text{s}$ the influence of the different time-step sizes is visible, all temperature profiles are in close agreement for a later time $t_2=50~\text{s}$. This is the case as most dynamics are happening in the beginning of the simulation as temperature differences are the largest and just starting to equilibrate. For the smallest time-step size shown, $\Delta t = 0.3125~\text{s}$, the results are converged in time.

\begin{figure}
        \begin{subfigure}{0.49\linewidth}
            \centering
            \includegraphics[width=\linewidth]{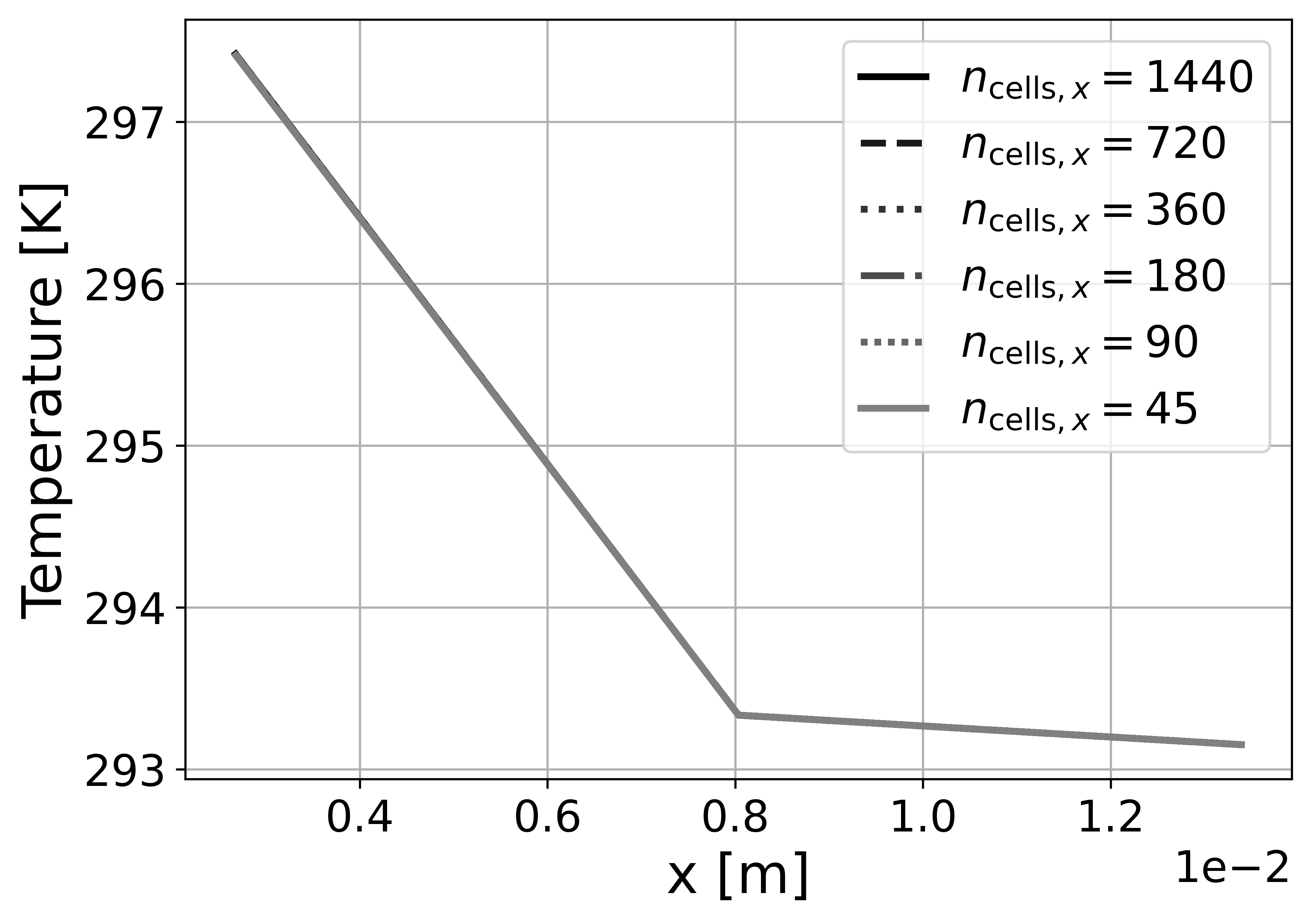}
            \caption{$t_1=5$s}
            \label{fig:rev_conv_space_lte_t1}
        \end{subfigure} \hfill
        \begin{subfigure}{0.49\linewidth}
            \centering
            \includegraphics[width=\linewidth]{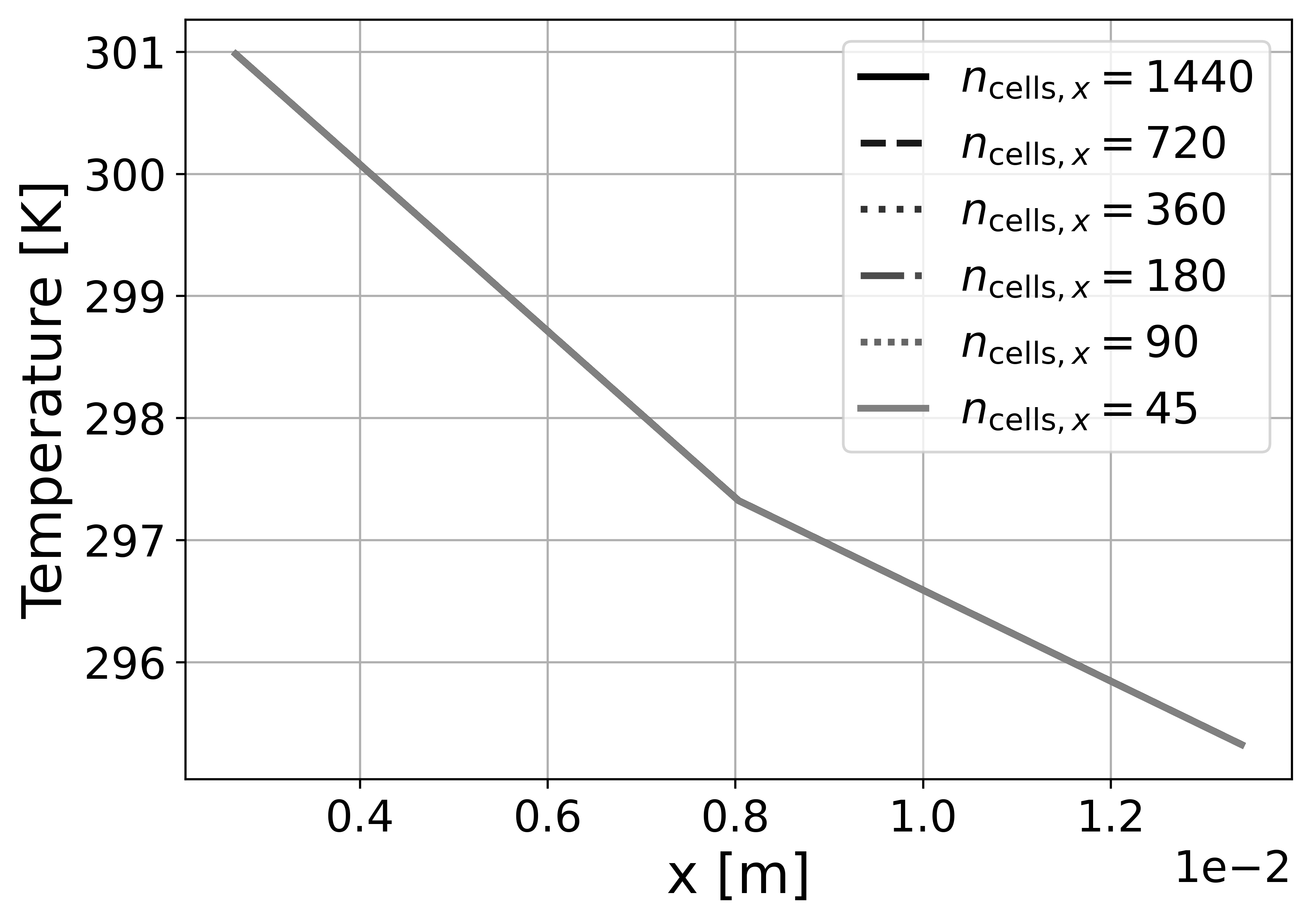}
            \caption{$t_2=50$s}
            \label{fig:rev_conv_space_lte_t2}
        \end{subfigure}
        \caption{Average temperatures for the REV-scale model in case of LTE for different numbers of discretization cells $n_{\mathrm{cells},x}$. On the left one earlier time at $t_1=5~\mathrm{s}$ is shown and on the right a later time $t_2=50~\mathrm{s}$. The time-step size is $\Delta t = 0.3125~\mathrm{s}$.}
        \label{fig:rev_conv_space_lte}
    \end{figure}

\begin{figure}
        \begin{subfigure}{\linewidth}
            \centering
            \includegraphics[width=\linewidth]{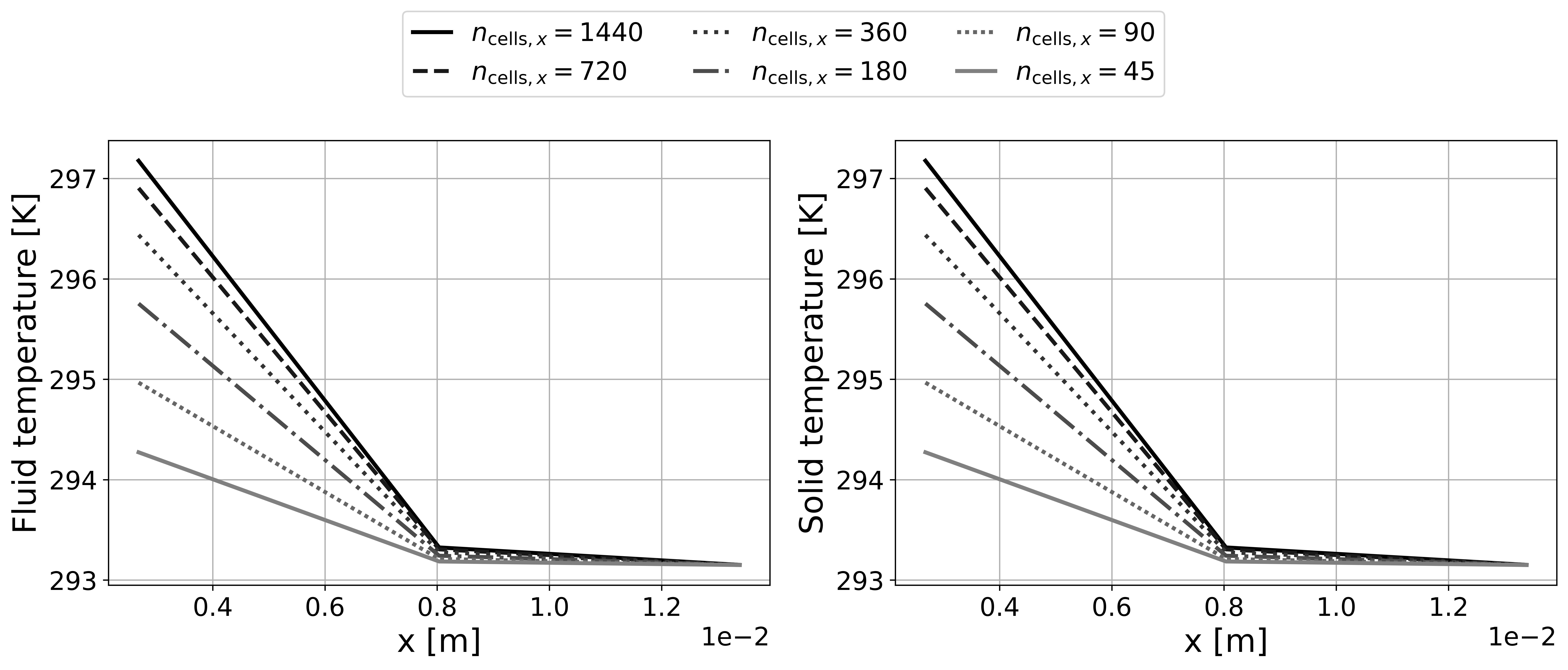}
            \caption{$t_1=5$s}
            \label{fig:rev_conv_space_ltne1_t1}
        \end{subfigure}
        \begin{subfigure}{\linewidth}
            \centering
            \includegraphics[width=\linewidth]{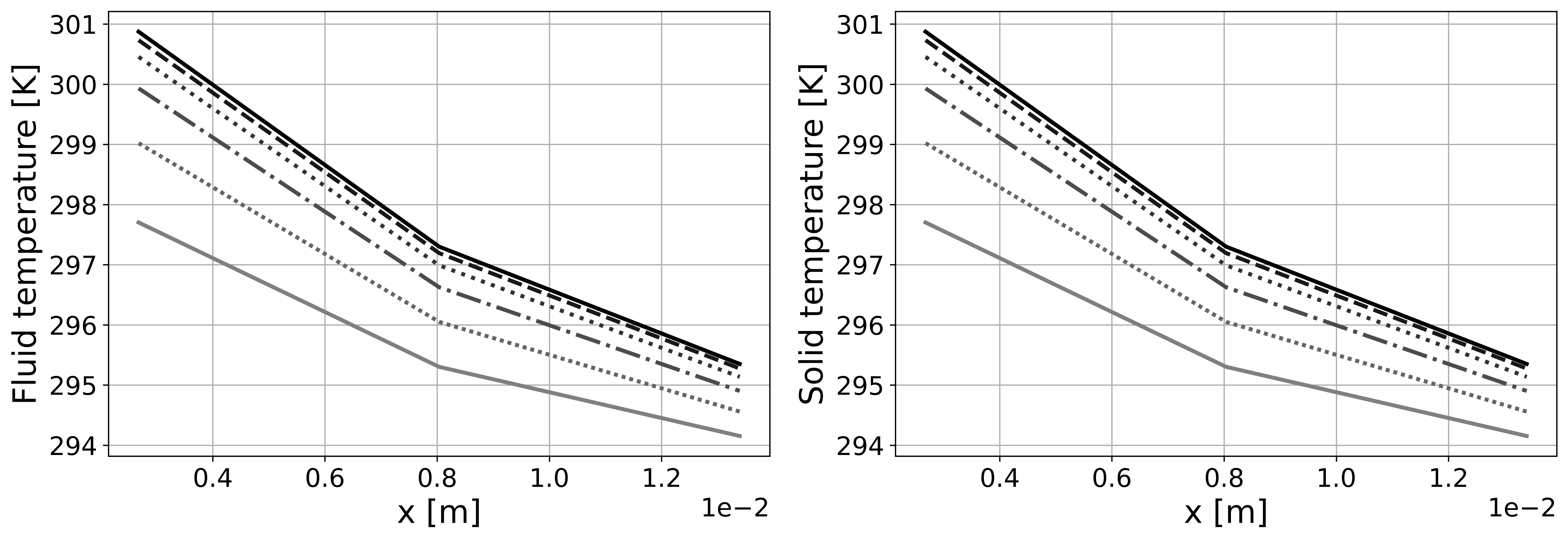}
            \caption{$t_2=50$s}
            \label{fig:rev_conv_space_ltne1_t2}
        \end{subfigure}
        \caption{Average temperatures for the REV-scale model in case of LTNE with $h=3.8\times10^7~\frac{\mathrm{W}}{\mathrm{m}^2\mathrm{K}}$ (as in \Cref{sec:case1}) for different numbers of discretization cells $n_{\mathrm{cells},x}$. Fluid and solid temperatures are shown on the left and right, while one earlier time with $t_1=5~\mathrm{s}$ is shown at the top row and one later time $t_2=50~\mathrm{s}$ at the bottom row. The time-step size is $\Delta t = 0.3125~\mathrm{s}$.}
        \label{fig:rev_conv_space_ltne1}
    \end{figure}

\begin{figure}
        \begin{subfigure}{\linewidth}
            \centering
            \includegraphics[width=\linewidth]{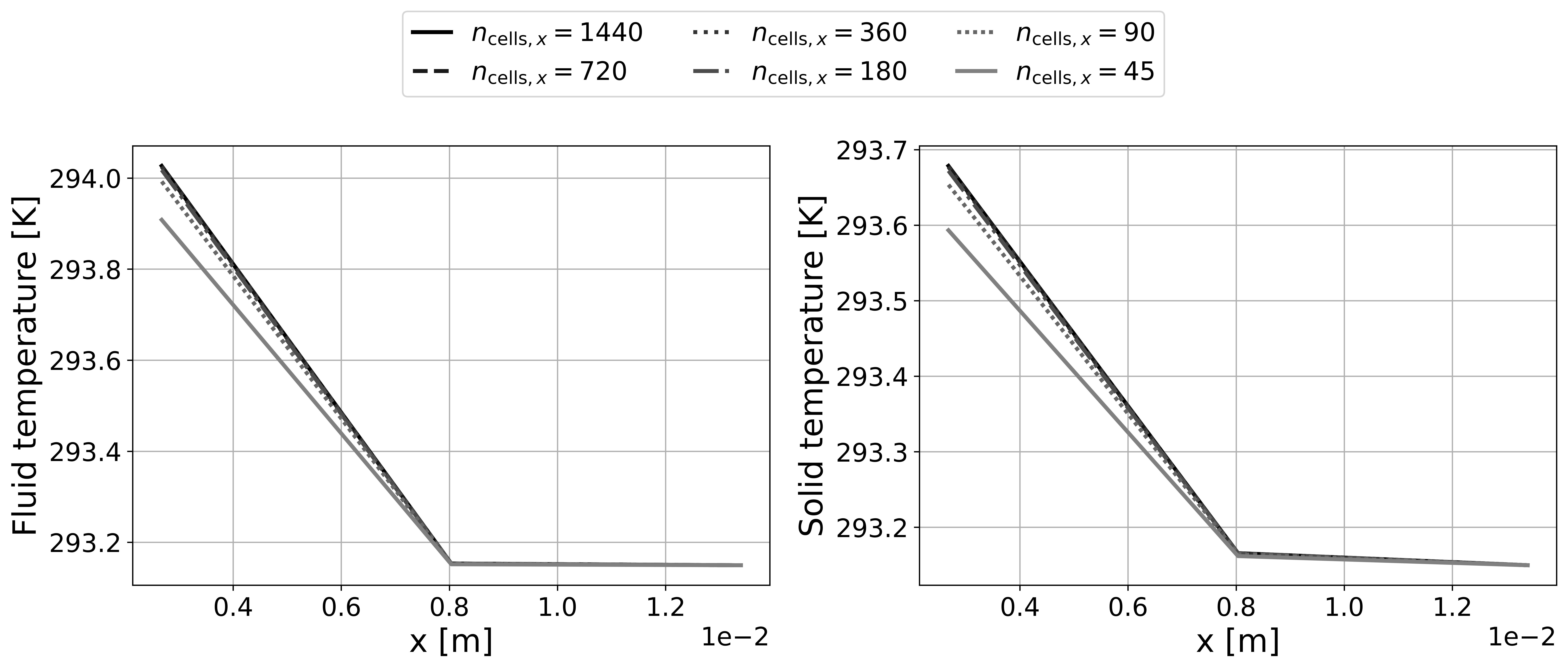}
            \caption{$t_1=5$s}
            \label{fig:rev_conv_space_ltne2_t1}
        \end{subfigure}
        \begin{subfigure}{\linewidth}
            \centering
            \includegraphics[width=\linewidth]{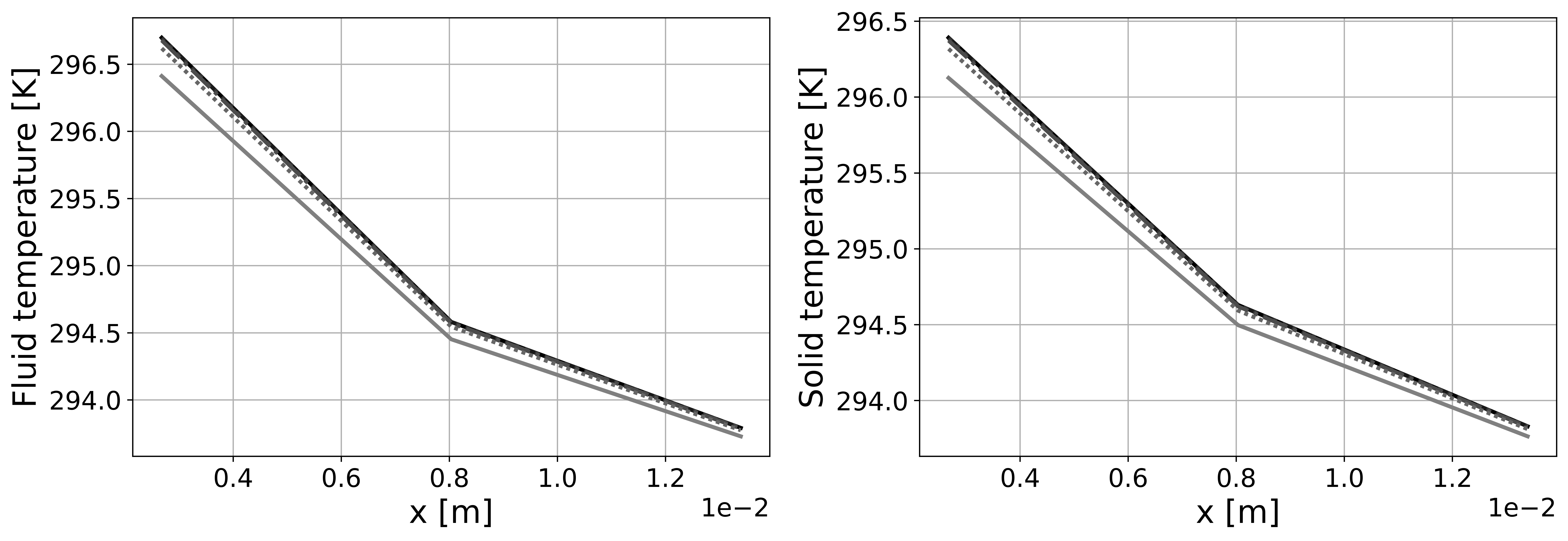}
            \caption{$t_2=50$s}
            \label{fig:rev_conv_space_ltne2_t2}
        \end{subfigure}
        \caption{Average temperatures for the REV-scale model in case of LTNE with $h=100~\frac{\mathrm{W}}{\mathrm{m}^2\mathrm{K}}$ (as in \Cref{sec:case2}) for different numbers of discretization cells $n_{\mathrm{cells},x}$. Fluid and solid temperatures are shown on the left and right, while one earlier time with $t_1=5~\mathrm{s}$ is shown at the top row and one later time $t_2=50~\mathrm{s}$ at the bottom row. The time-step size is $\Delta t = 0.3125~\mathrm{s}$.}
        \label{fig:rev_conv_space_ltne2}
    \end{figure}

    The figures \ref{fig:rev_conv_space_lte},  \ref{fig:rev_conv_space_ltne1} and \ref{fig:rev_conv_space_ltne2} show the influence of the number of discretization cells in the primary direction of the dynamics, the $x-$direction. While for the local equilibrium model, the discretization in space has no influence on the results (see \Cref{fig:rev_conv_space_lte}), the other two cases with the LTNE model show a dependence on the spatial discretization. For $h=100~\frac{\mathrm{W}}{\mathrm{m}^2\mathrm{K}}$, results for both phase temperatures are not visibly changing anymore for numbers of cells larger than $n_{\text{cells},x}=180$ (see \Cref{fig:rev_conv_space_ltne2}). However, for the case of $h=8.3\times 10^7 ~\frac{\mathrm{W}}{\mathrm{m}^2\mathrm{K}}$, a finer discretization in space is needed, especially close to the left boundary of the domain, where a higher fluid temperature is imposed (see \Cref{fig:rev_conv_space_ltne1}). Note that for smaller spatial discretizations with $n_{\text{cells},x}=1440$, the gradient of the resulting temperature profile, in particular at an earlier time, is larger compared to coarser spatial discretization as the region of high temperature differences are better resolved. We consider a discretization with $n_{\text{cells},x}=1440$ to be fine enough.
    
\newpage
\section{Influence of number of reference cells in \texorpdfstring{$y-$}{y} and \texorpdfstring{$z-$}{z} direction on average temperatures} \label{app:nRef_influence}
\begin{figure}
        \begin{subfigure}{\linewidth}
            \centering
            \includegraphics[width=\linewidth]{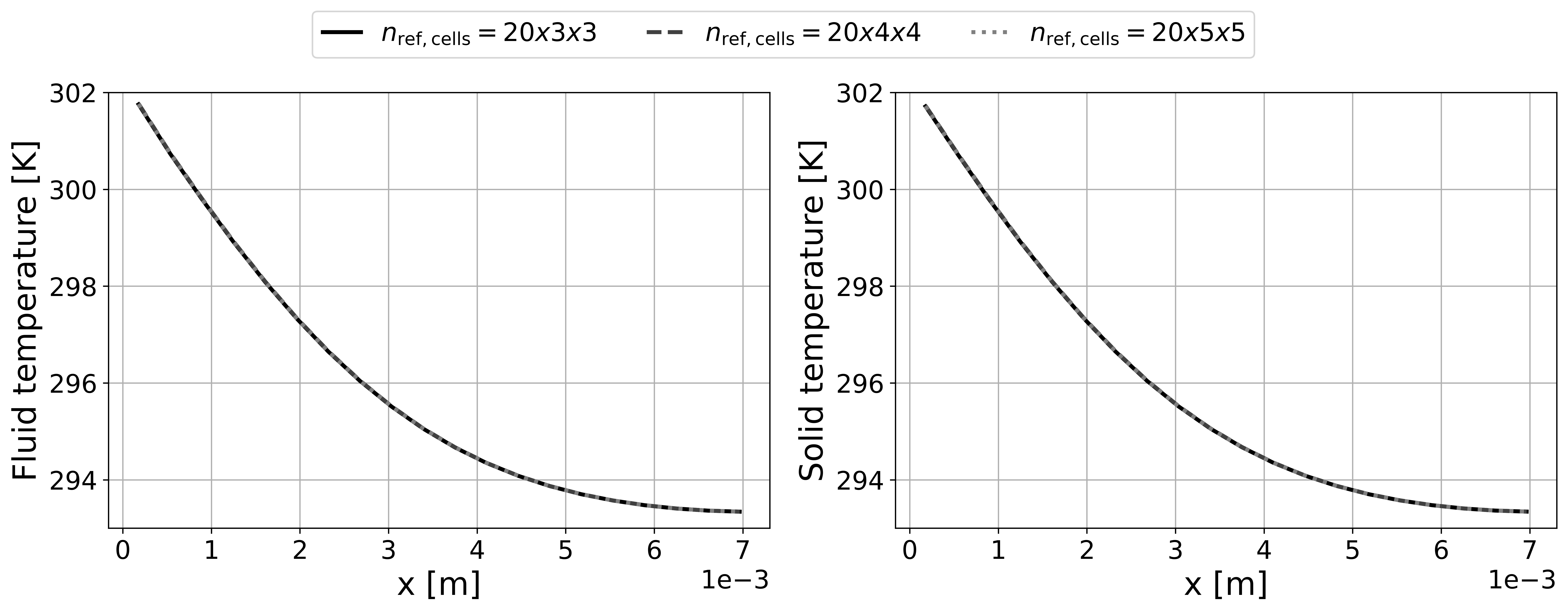}
            \caption{$t_1=5$s}
            \label{fig:pr_influence_yz_t1}
        \end{subfigure}
        \begin{subfigure}{\linewidth}
            \centering
            \includegraphics[width=\linewidth]{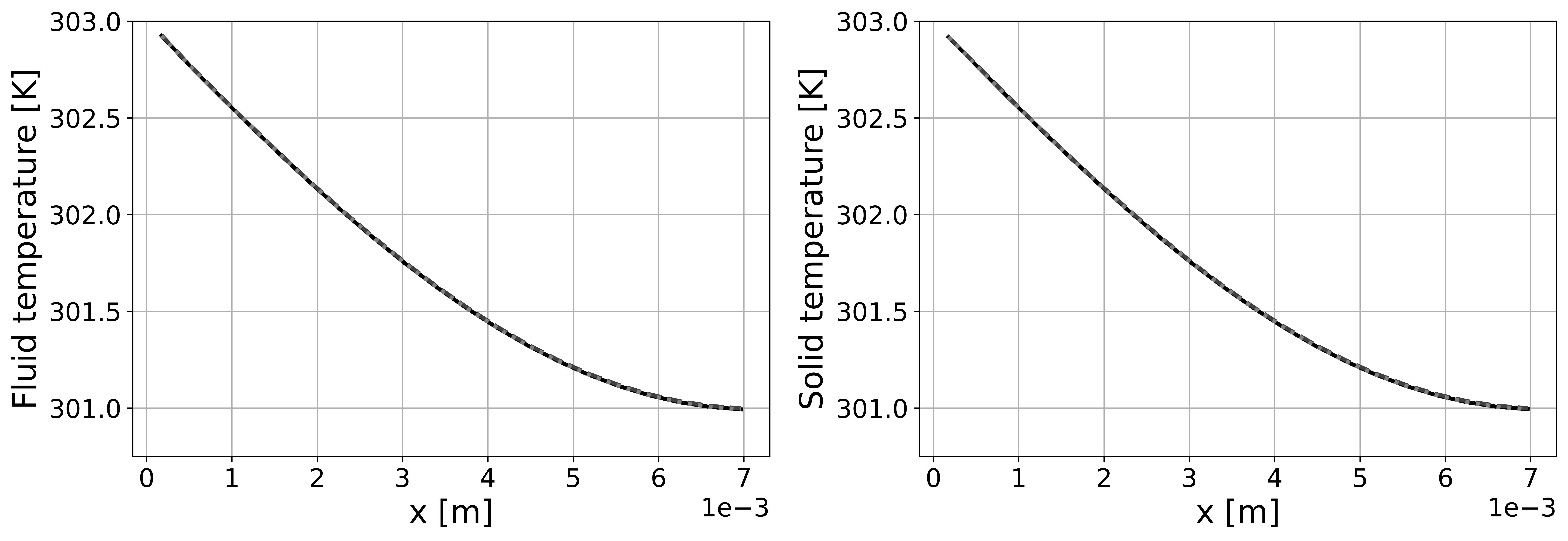}
            \caption{$t_2=50$s}
            \label{fig:pr_influence_yz_t2}
        \end{subfigure}
        \caption{Influence of number of reference cells in $y-$ and $z-$direction on the resulting temperature profile using a pore-resolved model with temperature continuity at the fluid-solid interface.}
        \label{fig:pr_influence_yz}
    \end{figure}

    In \Cref{fig:pr_influence_yz}, the influence of different numbers of reference cells in $y-$ and $z-$direction are investigated while holding $n_{\mathrm{cells},x}^{\mathrm{ref}}$ constant. The pore-resolved model is chosen for this investigation as it gives the most detailed results of all three model classes. Compared to \Cref{sec:comparison}, the number of reference cells in $x-$direction is decreased to $n_{\mathrm{cells},x}^{\mathrm{ref}}=20$ such that more cells in $y-$ and $z-$ direction can also be investigated. Temperatures are averaged over $1\times n_{\mathrm{cells},y}^{\mathrm{ref}}\times n_{\mathrm{cells},z}^{\mathrm{ref}}$ reference cells. As we see in \Cref{fig:pr_influence_yz} that all temperature profiles for different $n_{\mathrm{cells},y}^{\mathrm{ref}}$ and $ n_{\mathrm{cells},z}^{\mathrm{ref}}$ are identical, this influence can be neglected. This is the case as the boundary conditions of the setup (see \Cref{sec:comparison}) were chosen such that all the boundaries feature zero gradient boundary conditions for the temperature except on the left boundary, where a higher temperature is fixed for the fluid phase.

\newpage
\section{Comparison of dual-network model to REV-scale model with coarse spatial discretization} \label{app:coarseREV}
\begin{figure}
        \begin{subfigure}{\linewidth}
            \centering
            \includegraphics[width=\linewidth]{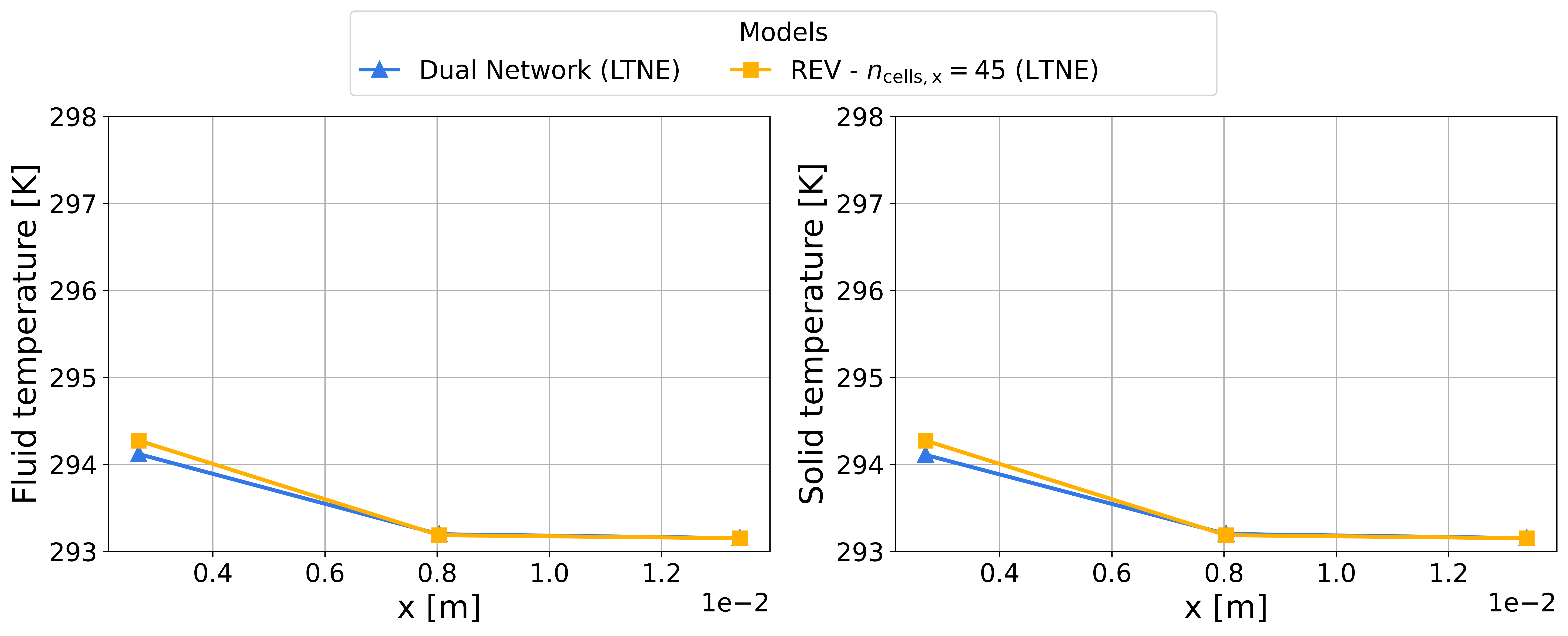}
            \caption{$t=5$s}
            \label{fig:case1_t1_coarse}
        \end{subfigure}
        \begin{subfigure}{\linewidth}
            \centering
            \includegraphics[width=\linewidth]{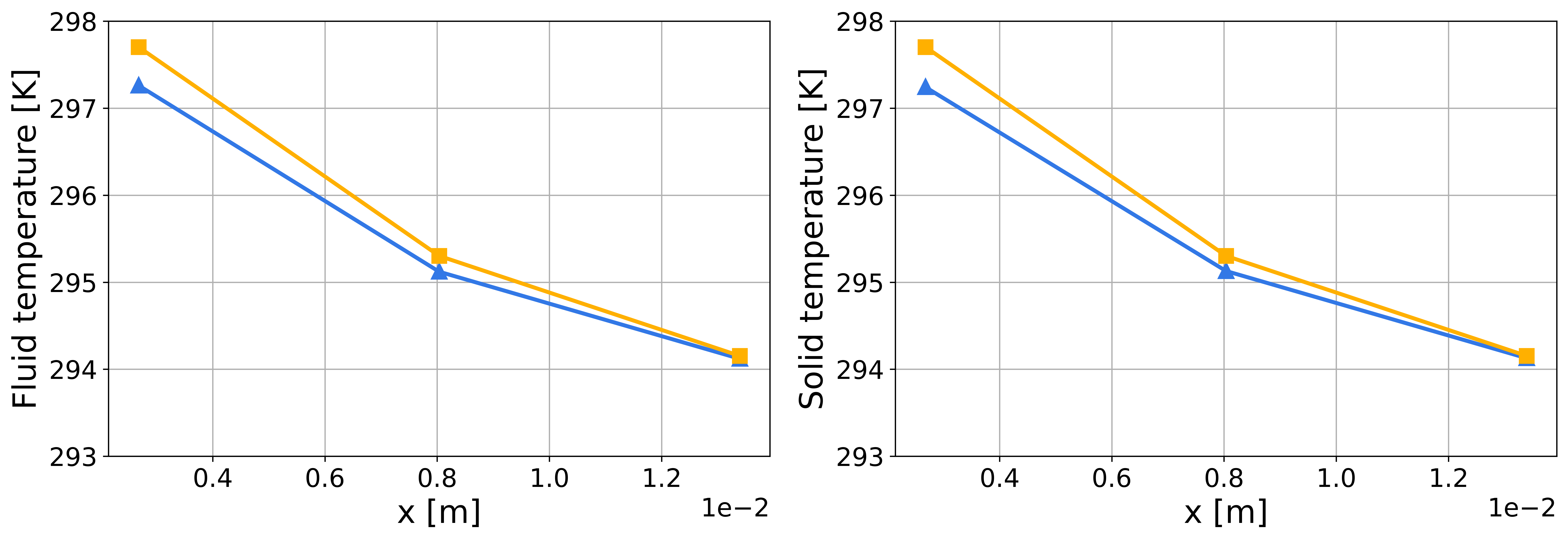}
            \caption{$t=50$s}
            \label{fig:case1_t2_coarse}
        \end{subfigure}
        \caption{Averaged REV-scale temperatures for dual-network model and REV-scale model. The models account for LTNE concepts, where the interfacial heat transfer coefficient is $h_1 = 8.3\times10^7 ~\frac{\text{W}}{\text{m}^2\text{K}}$. Results are shown for fluid (left) and solid temperatures (right) for an earlier time $t_1$ (top) and a later time $t_2$ (bottom).}
        \label{fig:case1_coarse}
    \end{figure}

\begin{figure}
        \begin{subfigure}{\linewidth}
            \centering
            \includegraphics[width=\linewidth]{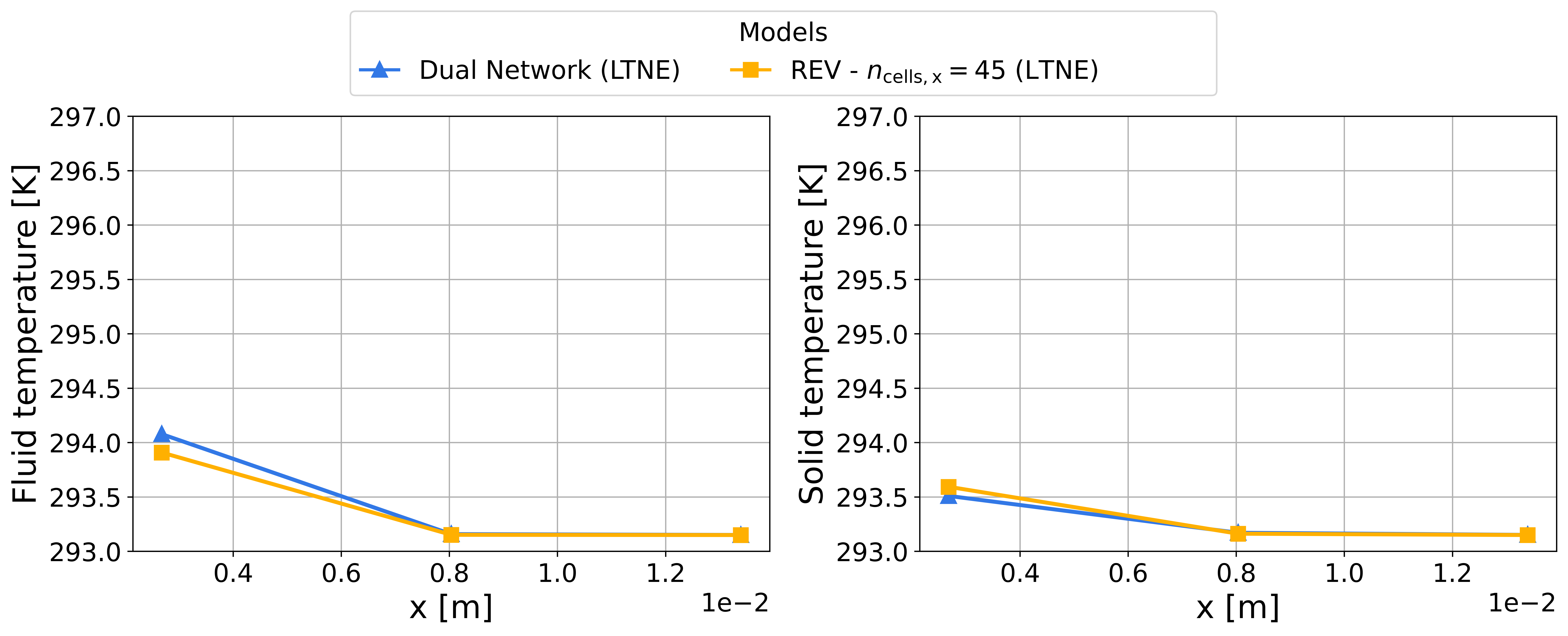}
            \caption{$t=5$s}
            \label{fig:case2_t1_coarse}
        \end{subfigure}
        \begin{subfigure}{\linewidth}
            \centering
            \includegraphics[width=\linewidth]{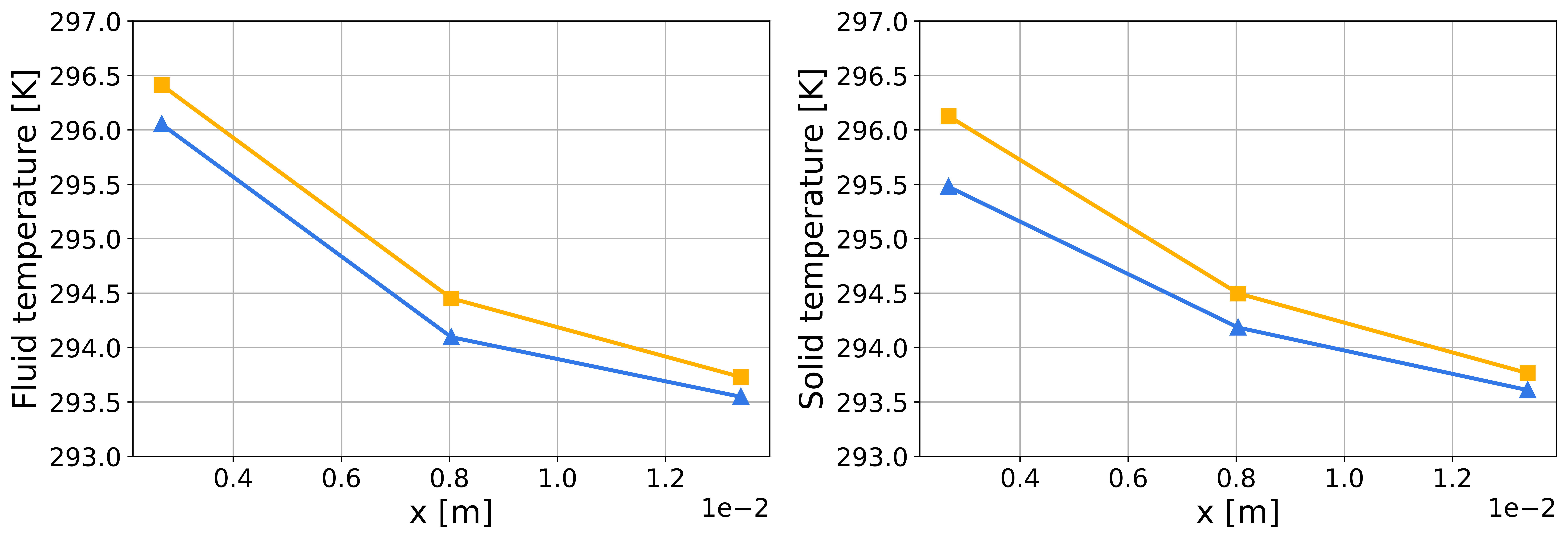}
            \caption{$t=50$s}
            \label{fig:case2_t2_coarse}
        \end{subfigure}
        \caption{Averaged REV-scale temperatures for dual-network model and REV-scale model. The models account for LTNE concepts, where the interfacial heat transfer coefficient is $h_2 = 100 ~\frac{\text{W}}{\text{m}^2\text{K}}$. Results are shown for fluid (left) and solid temperatures (right) for an earlier time $t_1$ (top) and a later time $t_2$ (bottom).}
        \label{fig:case2_coarse}
    \end{figure}

    In \Cref{fig:case1_coarse} and \Cref{fig:case2_coarse} the dual-network model results are shown against results of REV simulation with a coarse spatial resolution for $h_1=8.3\times10^7 ~\frac{\text{W}}{\text{m}^2\text{K}}$ and $h_2 = 100 ~\frac{\text{W}}{\text{m}^2\text{K}}$ respectively. The spatial discretization of the REV model is chosen to be comparable to the resolution of the dual-network model and hence $n_{\mathrm{cell,x}}=45$ cells are chosen. Note here that $45$ cells relates to the number of reference cells in $x-$direction in the domain. The resulting temperatures for those two cases are much closer in comparison to fine resolution for the REV model (cf. \Cref{sec:comparison}), indicating the influence of the spatial discretization.

\newpage
\section*{Acknowlegements}
We acknowledge funding by the Deutsche Forschungsgemeinschaft (DFG, German Research Foundation), within the framework of the Collaborative Research Centre “Interface-Driven Multi-Field Processes in Porous Media” (SFB 1313, Project Number 327154368), the International Research Training Group "Droplet Interaction Technologies" (GRK 2160: DROPIT, Project Number 270852890), the VISTA program, "The Norwegian Academy of Science and Letters and Equinor" and the Research Council of Norway, Equinor ASA, A/S Norske Shell, Harbour Energy Norge AS through project ExpReCCS (grant number 336294).
    
\section*{Funding}
Funded by the Deutsche Forschungsgemeinschaft (DFG, German Research Foundation) through funding the Collaborative Research Center on "Interface-Driven Multi-Field Processes in Porous Media" (SFB 1313, Project Number 327154368) and through funding the International Research Training Group "Droplet Interaction Technologies" (GRK 2160/1: DROPIT, Project Number 270852890).
Furthermore, funded by "The Norwegian Academy of Science and Letters and Equinor” through the VISTA program and funded by Research Council of Norway, Equinor ASA, A/S Norske Shell, Harbour Energy Norge AS through project ExpReCCS (grant number 336294).
Open Access funding enabled and organized by Project DEAL.

\section*{Availability of code and data}
    The source codes for the simulations performed with \DuMuX,  PorePy and Netgen/NGSolve are available at \url{https://doi.org/10.18419/DARUS-4781} \citep{kostelecky2025code}, \url{https://doi.org/10.18419/DARUS-4785} \citep{stefansson2025code} and  \url{https://doi.org/10.18419/DARUS-4771}  \citep{bringedal2025code} respectively. Additionally, scripts for post-processing the results are provided in \url{https://doi.org/10.18419/DARUS-4782} \citep{kostelecky2025postprocessing}.

\section*{CRediT authorship contribution statement}
\textbf{Anna Mareike Kostelecky}: Conceptualization, Data Curation, Investigation, Methodology, Software, Validation, Visualization, Writing – original draft, Writing – review \& editing. \textbf{Ivar Stefansson}: Conceptualization, Data Curation, Investigation, Methodology, Software, Validation, Writing – review \& editing. \textbf{Carina Bringedal}: Conceptualization, Data Curation, Funding acquisition, Investigation, Methodology, Software, Validation, Writing – original draft, Writing – review \& editing. \textbf{Tufan Ghosh}: Conceptualization, Funding acquisition, Investigation, Methodology, Software, Writing – review \& editing. \textbf{Helge K. Dahle}: Conceptualization, Funding acquisition, Methodology, Writing – review \& editing. \textbf{Rainer Helmig} Conceptualization, Funding acquisition, Methodology, Supervision, Writing – review \& editing.


\bibliographystyle{elsarticle-harv} 
\bibliography{bibliography}

\end{document}